\documentclass[10pt,preprint]{aastex}

\usepackage{natbib,psfig}
\bibpunct{(}{)}{;}{a}{}{,}

\newcommand{\asca}{{\emph{ASCA}}}
\newcommand{\chandra}{{\emph{Chandra}}}
\newcommand{\xmm}{\emph{XMM-Newton}}

\newcommand{\rosat}{{\emph{ROSAT}}}

\newcommand{\kms}{\mbox{\,km\,s$^{-1}$}}
\newcommand{\cmsq}{\mbox{\,cm$^{-2}$}}

\newcommand{\flux}{\mbox{\,ergs~cm$^{-2}$~s$^{-1}$}}

\newcommand{\fE}{\mbox{\,photons~cm$^{-2}$~s$^{-1}$~keV$^{-1}$}}
\newcommand{\fnu}{\mbox{\,ergs~cm$^{-2}$~s$^{-1}$~Hz$^{-1}$}}
\newcommand{\lumin}{\mbox{\,ergs~s$^{-1}$}}
\newcommand{\persec}{\mbox{\,s$^{-1}$}}
\newcommand{\nh}{\mbox{${N}_{\rm H}$}} 

\newcommand{\aox}{$\alpha_{\rm ox}$}

\newcommand{\alphauv}{$\alpha_{\rm uv}$}

\newcommand{\CIV}{\ion{C}{4}}
\newcommand{\MgII}{\ion{Mg}{2}}
\newcommand{\OIII}{[\ion{O}{3}]}
\newcommand{\HI}{\ion{H}{1}}
\newcommand{\delg}{$\Delta(g-i)$}
\newcommand{\daox}{$\Delta \alpha_{\rm ox}$}
\newcommand{\delv}{$\Delta v_{\rm b}$}
\newcommand{\HR}{{HR}}
\newcommand{\GHR}{$\Gamma_{\rm HR}$}

\begin{document}

\received{}
\accepted{}
\slugcomment{\date}

\title{X-ray Insights Into Interpreting \CIV\ Blueshifts and Optical/UV Continua}

\author{
S.\ C. Gallagher,\altaffilmark{1}
Gordon T. Richards,\altaffilmark{2}
Patrick B. Hall,\altaffilmark{2,3}
W.\ N. Brandt,\altaffilmark{4}
Donald P. Schneider,\altaffilmark{4}
Daniel E. Vanden Berk\altaffilmark{5}
}
\altaffiltext{1}{Division of Astronomy and Astrophysics, University of
  California, Los Angeles, 405 Hilgard Avenue, Los Angeles, CA 90095.}
\altaffiltext{2}{Princeton University Observatory, Princeton, NJ 08544.}
\altaffiltext{3}{Department of Physics \& Astronomy, York University,
  4700 Keele Street, Toronto, ON M3J 1P3, Canada.}
\altaffiltext{4}{Department of Astronomy and Astrophysics, The
  Pennsylvania State University, University Park, PA 16802.}
\altaffiltext{5}{Department of Physics and Astronomy, University of
  Pittsburgh, 3941 O'Hara Street, Pittsburgh, PA 15260.}

\begin{abstract}
We present 0.5--8.0~keV 
\chandra\ observations of six bright quasars that represent
extrema in quasar emission-line properties --- three quasars each 
with small and large blueshifts of the \CIV\ emission
line with respect to the systemic redshift of the quasars.
Supplemented with seven archival \chandra\ observations of quasars that met our
selection criteria, we investigate the origin of this emission-line
phenomenon in the general context of the structure of quasars.  
We find that the quasars with the largest \CIV\ blueshifts show
evidence, from joint-spectral fitting, for intrinsic X-ray
absorption (\nh$\sim10^{22}$\cmsq).  Given the lack of accompanying
\CIV\ absorption, this gas is likely to be highly ionized, and may be
identified with the shielding gas in the disk-wind paradigm. 
Furthermore, we find evidence for a correlation
of \alphauv, the ultraviolet spectral index,
with the hardness of the X-ray continuum; an analysis of independent
Bright Quasar Survey data from the literature supports this conclusion. 
This result points to intrinsically red quasars having systematically
flatter hard X-ray continua without evidence for X-ray absorption. 
We speculate on the origins of these correlations of X-ray properties
with both \CIV\ blueshift and \alphauv\ and discuss the implications for
models of quasar structure.  
\end{abstract}


\section{Introduction}

Considering the dynamic range of black hole masses and luminosities,
quasar emission-line phenomenology is remarkably consistent.  At the same
time, the physical drivers for the known range of emission-line
properties are poorly understood.

One of the reasons the emission-line region is not better
understood is because of the general similarity of quasar spectra
\citep{rfs+01} and the lack of sufficient emission-line distinctions to test
competing models.  That said, there are a few well-known
differences.  For example,
it is generally believed that the widths of the Balmer lines (and
probably \ion{Mg}{2}) are related to the mass of a quasar's central
black hole \citep[e.g.,][]{ves02} --- although the presumed disk-like
configuration of this gas means that the line widths will be affected
by projection effects \citep[e.g.,][]{kro01}.  There is also the observation
that more luminous quasars tend to have \CIV\ emission lines with
smaller equivalent widths, otherwise known as the \citet{bal77}
effect.  Recently, \citet{bl04} have argued that the Baldwin effect
and the dynamics of the broad emission-line region (BELR) in general are
driven by differences in the Eddington ratio, $L/L_{\rm
  Edd}$, among quasars.  Similarly,
\citet{bg92} and others have proposed that $L/L_{\rm Edd}$ drives a third
trend, mainly the anti-correlation between the strengths of \OIII\
and \ion{Fe}{2} emission in the optical part of the spectrum.
A fourth effect seen in quasar emission lines is 
emission-line blueshifting \citep[e.g.,][]{gas82,wil84,tf92}, where higher
ionization lines yield emission-line redshifts that are
systematically too small --- as if the lines had been shifted
blueward.

\citet{ric+02} recently presented a summary of the emission-line
blueshift\footnote{The blueshift is defined as the velocity offset between redshifts
derived separately from \CIV\ and \MgII, given in velocity units with
more positive velocity indicating larger emission-line blueshifts.}
effect among a sample of 3814 quasars from the Sloan Digital
Sky Survey (SDSS; \citealt{yor+00}).  They showed that these
emission-line shifts are more ubiquitous than previously thought, with
the average quasar having a \CIV\ emission line that yields a redshift
that is too small with respect to \MgII\ by \mbox{$\sim800$\kms}.
Furthermore, quasar composite spectra created from samples binned
by \CIV\ blueshift showed differences in their ultraviolet
continua and other emission-line properties.  Specifically, the
composite spectrum of the largest-blueshift quasars had a significantly
bluer than average ultraviolet continuum and weaker \CIV\ emission,
while the smallest-blueshift composite was notably redder with
stronger \CIV\ emission.  \citet{ric+02} showed that this difference
resulted  from the intrinsic continuum color of the quasars, rather than
as an effect of reddening.

The distribution of \CIV\ blueshifts ranges over 3000\kms, making this effect an excellent
candidate for further investigation into the structure of the BELR.
\citet{ric+02} hypothesized that the range of \CIV\ blueshift and
ultraviolet continuum color was a consequence of orientation.
Orientation could be understood in one of two ways, either with
respect to the accretion disk, or with respect to the wind, which may
have a range of opening angles (for example, see Figure~7 in
\citealt{elv00} and Figure~5 in \citealt{ric+04}).  
In both cases, understanding the BELR might be aided by probing
smaller scales.
X-ray emission is believed to arise on physical scales comparable to and
smaller than the optical-UV continuum. Therefore, X-rays are not only
sensitive to physical conditions in the immediate vicinity of the black hole, 
they can also probe the line of sight to the observer from the 
inner regions of the accretion flow. 
X-ray studies thus offer the potential to tie BELR phenomena (on
scales of light days) to fundamental properties of the accretion
disk.  For example, if the range of BELR phenomenology results purely from 
orientation effects, one might expect the intrinsic X-ray continuum
properties to be unrelated.  

To investigate this avenue,
we undertook an exploratory \chandra\ \citep{chandra_ref} survey to study the X-ray
properties of extreme examples of the \CIV\ blueshift distribution.  
We were awarded six observations in Cycle~4; these observations have
been supplemented with seven from the archive. 
In \S\ref{sec:data}, we present the target selection process and describe 
the initial \chandra\ data analysis.  We study the trends
in these data in \S\ref{sec:trends}.   The results from
joint spectral fitting of the \chandra\ spectra and the connection
to optical/UV emission-line and continuum properties are presented 
in \S\ref{sec:joint}.
In \S\ref{sec:s}, we summarize the results from our X-ray analysis, and introduce
comparison data from the literature on Bright Quasar Survey \citep[BQS;][]{bqs} objects.
In \S\ref{sec:d}, we discuss the possible physical origins of our
observed correlations and comment on the relative importance of external
orientation (tilt of the accretion disk with respect to our line of
sight), $L/L_{\rm Edd}$, and internal orientation
effects (such as the disk-wind opening angle --- possibly driven by
$L/L_{\rm Edd}$).  The final section briefly outlines our conclusions.

The cosmology assumed throughout has $H_{\rm 0}=70$\kms\,Mpc$^{-1}$,
$\Omega_{\rm m}=0.3$, and $\Omega_{\Lambda}=0.7$.

\section{X-ray Observations and Data Analysis}
\label{sec:data}

\subsection{Sample Selection}

Our initial sample was selected from the 3814 Early Data Release
(EDR) Quasars \citep{sch+02}.  The sample was first restricted to the 794 quasars
with $1.54<z<2.20$ to enable measurement of redshifts from both \CIV\
and \MgII\ broad emission lines in the SDSS spectra.  The
velocity offsets between the \CIV\ and \MgII\ emission-line redshifts
were computed, and the quasars were divided into four broad blueshift
bins from small to large \CIV\ blueshift. 
From the smallest and the largest blueshift bins, we
requested time for exploratory \chandra\ Advanced CCD Imaging Spectrometer
\citep[ACIS;][]{acis_ref} observations, giving priority to the
optically brightest quasars with small Galactic \nh.
The sample was vetted to exclude broad absorption line (BAL) quasars and quasars detected in the 20~cm FIRST
survey \citep{bwh95}. In Cycle~4 we were awarded time to observe six
quasars for this program.

In addition to the six targets in our primary
program, we also included seven additional targets from the \chandra\
archive to increase the sample size.  These archival data were found
by cross-correlating the complete list of SDSS Data Release 1 (DR1)
quasars \citep{sch+03} with the same redshift range as the primary
sample with observations publicly available from the
\chandra\ ACIS archive as
of March 2004. Of these, only J2348$+$0057\footnote{This
quasar is a member of a wide quasar binary (Q2345$+$007AB;
\citealt{weedman+82}); the fainter member of the pair is not included in
our study as it has no SDSS spectrum.} was an observation target \citep{gre+02}; the rest were
serendipitous.  The target list and optical
properties of all of the quasars in this program 
are presented in Table~\ref{tab:opt}.  The optical properties
include \delv\ (the \CIV\ blueshift), \delg\ (the observed $g-i$ minus the
median of the DR1 $g-i$ distribution at the redshift of the quasar;
\citealt{rfs+01}), and \alphauv\ (the spectral index of a power-law
continuum fit to the optical spectra between rest-frame 1450 and 2200 \AA).
The optical spectra are shown in Figures~\ref{fig:gtr1}--\ref{fig:gtr2}.   
A catalog of the \chandra\ observations is presented in Table~\ref{tab:log}.
The additional archival targets are listed below the
primary sample in the data tables.

All archival targets were checked to exclude
radio-detected and BAL quasars.  The archival quasar, J0200$-$0845,
was omitted from subsequent analysis because it is a radio-loud
BAL~quasar. We include it in the data tables
to present the X-ray data for the record.  

\subsection{X-ray Analysis}

\chandra\ observed the six extreme \CIV\ blueshift targets (three each of
small and large blueshifts) between 2002 November 20 and 2003 August
26. Each target was observed at the aimpoint of the back-illuminated S3 CCD of
ACIS in faint mode with exposure times ranging from 3.5 to 5.6~ks.
The data were processed using the standard \chandra\ X-ray Center
(CXC) aspect solution and grade filtering from which the level 2
events file was taken.  

Both aperture photometry and the CIAO 3.0\footnote{http://cxc.harvard.edu/ciao} 
wavelet detection tool {\em
wavdetect} \citep{wavdet_ref} were used in the soft (0.5--2.0~keV), hard (2.0--8.0~keV),
and full (0.5--8.0~keV) bands to determine the measured counts for a
point source in each band.  The lower limit of 0.5~keV was chosen to
match the well-calibrated part of the response; above 8.0~keV, the
effective area of the \chandra\ mirrors drops considerably and the
background increases significantly.  A 60$\arcsec\times60\arcsec$
image (14400 pixels) around the known optical position of each quasar
was searched with {\em wavdetect} at the 4$\times10^{-5}$
false-probability threshold for the full, soft, and hard bands.
Wavelet scale
sizes were 1, 1.414, 2, 2.828, 4, and 5.66 pixels. For aperture photometry of
the on-axis targets, the counts were extracted from circular source
cells centered on the SDSS optical positions with a 2$\farcs$5 radius.
For the off-axis targets, a circular or elliptical source aperture was
used.  The aperture size and shape were chosen to include at least
$90\%$ of the encircled energy at 6.4~keV (from Fig.~4.10 of the
\chandra\ Proposers' Observatory Guide).  For the new observations,
the quasars were each significantly detected with full-band counts
ranging from 27 to 89; the difference in counts between {\em
wavdetect} and aperture photometry was in all cases $\le1$~count.  
On average, the {\em wavdetect} centroid positions were
$\sim0\farcs26$ from the SDSS optical positions; the largest offset was
$0\farcs53$ for the entire dataset.  For the mean flux of our sample
($\sim7\times10^{-14}$\flux), the source density from blank field
number counts is (1.5--3.9)$\times10^{-6}$ sources arcsec$^{-2}$ \citep[e.g.,][]{xrb}, and
so the chance of a misidentification is negligible.
  
For the off-axis archival observations, the {\em wavdetect}
photometry was found to be unreliable, particularly in the hard band,
and so aperture photometry was used for the subsequent calculations
for these quasars.  For the on-axis observations, the background was
in all cases negligible ($<$1 count in the source region).  For the
off-axis observations, the source counts are
background-subtracted. The background was determined from a
source-free elliptical or circular annulus around the source aperture
and normalized to the area of the source region.  In all cases
backgrounds were still low; the largest background contributed
$\lesssim6\%$ of the net full-band counts within the source aperture
(for J1204$+$0150).

For each target, to provide a coarse quantitative measure of the spectral 
shape we calculated the hardness ratio, defined as \HR$=(h-s)/(h+s)$, where 
$h$ and $s$ refer to the hard- and soft-band counts, respectively.
A typical, radio-quiet quasar has a power-law continuum in the 0.5--10.0~keV band
characterized by the photon index, $\Gamma$, and the 1~keV
normalization, $N_{\rm 1 keV}$: $f_{\rm E}=N_{\rm 1 keV}E^{-\Gamma}$
\fE.  From spectral fitting of \asca\ and other data, $\Gamma$ is found to
average $2.0\pm0.25$ for radio-quiet quasars
\citep[e.g.,][]{GeoEtal2000,ReTu2000}.  To transform the observed \HR\
into $\Gamma$, the X-ray spectral modeling tool {\em XSPEC}
\citep{xspec_ref} was used to simulate the response of the
instrument. The simulation procedures followed were slightly
different for the new and archival observations because the archival
data were taken from different CCDs requiring a unique response for
each observation.  

For each new target, the appropriate Galactic column density (see
Table~\ref{tab:opt}) and modified auxiliary response file (arf)
were included in the model.  The arf contains the
energy-dependent effective area and quantum efficiency of the
telescope, filter, and detector system. Each arf was modified using the tool {\em
contamarf}{\footnote{http://space.mit.edu/CXC/analysis/ACIS\_Contam/ACIS\_Contam.html}}
to take into account the time-dependent degradation of the ACIS low-energy effective area likely due to the accumulation of a layer of
hydrocarbon contaminant on the optical blocking filters \citep{acis_contam}.  
The arf is modified by multiplying the original by a time- and
energy-dependent function derived from the empirically determined optical
depth of the contaminant. The redistribution
matrix file (rmf) maps the energy of an incident X-ray into the space
of the observed charge distribution of the detector.  The arf and rmf,
both required to simulate the instrument response, were generated
using the CIAO 3.0 tools {\em mkarf} and {\em mkrmf}.  The detector
response to incident power-law spectra with varying $\Gamma$ was then
simulated.  The hardness ratio and errors were compared to the modeled
\HR\ to determine the \GHR\ (with errors) that would generate the
observed \HR.  For reference, a typical radio-quiet quasar with no
intrinsic absorption and a photon index, $\Gamma=2.0$, would be
observed to have \HR\,$\approx-0.63$ on axis.  The modeled full-band count rate
was normalized to the observed full-band count rate to obtain the
power-law normalization, $N_{\rm 1 keV}$.  With $N_{\rm 1 keV}$ and
\GHR, the 0.5--8.0~keV flux, $F_{\rm X}$, and the flux density at
rest-frame 2~keV, $f_{\rm 2 keV}$, were calculated.  The errors quoted for
these two values are the Poisson errors \citep{Gehrels} from the
full-band counts.   
Lastly, \aox=0.384\,$\log(f_{\rm 2 keV}/f_{\rm 2500})$         
and its associated uncertainties were calculated, where the factor 
0.384 is the logarithm of the ratio of the frequencies at which the flux 
densities are measured and $f_{\rm 2500}$ is the
average flux density within the rest-frame range $2500\pm25$\,\AA\ in the SDSS
optical spectrum.
For reference, the SDSS spectrophotometric calibration has a
typical uncertainty of $4\%$ (root-mean square) in the $r$ band \citep{sdss_dr2}.
The uncertainty on \aox\ is given by
$\sigma_{\alpha_{\rm ox}} = 0.4084 \sqrt{ 
        (\sigma_{f_{\rm 2keV}}/f_{\rm 2keV})^2
        + (\sigma_{f_{\rm 2500}}/f_{\rm 2500})^2}$, 
where $\sigma_{f_{\rm 2keV}}$ is just the Poisson uncertainty,\footnote{Using
the ACIS archival observations of $>15$~ks exposure time, we have verified
that X-ray variability on $\sim5$~ks timescales does not contribute any
significant excess variance to the measured X-ray flux.}
but $\sigma_{f_{\rm 2500}}$ includes the estimated effects of variability.
We use a formula from \citet{ivezic+04} to
calculate the characteristic variation in magnitudes at 2500\,\AA\
for each quasar's absolute magnitude, $M_i$, and for the rest-frame $\Delta t$
(in days) between the SDSS spectral and X-ray imaging epochs:
$\sigma_{m_{2500}} = (1 + 0.024 M_i) (\Delta t/2500)^{0.3}$,
which is then converted to the flux uncertainty $\sigma_{f_{\rm 2500}}$ and
used to calculate $\sigma_{\alpha_{\rm ox}}$.

For the archival data, the same procedure was followed with the
exception that the arfs and rmfs were generated for each observation
using the CIAO 3.0 script {\em psextract}.  This process also takes
into account the time-dependent change in the arfs. The source
extraction regions were the same as the apertures used for aperture
photometry. The X-ray properties and the MJDs of the SDSS optical
spectra used are all presented in Table~\ref{tab:xcalc}.

\subsection{Notes on UV Absorption and Individual Quasars}
\label{sec:notes}
Some of our objects have intervening or associated ultraviolet
absorbers.  If the former are damped Ly$\alpha$ absorbers (DLAs), they
may produce some appreciable X-ray absorption.
We use the criteria of \citet{rt00} to determine which intervening
absorbers have a 50$\%$ chance of being a DLA with \nh$>2\times10^{20}$\cmsq.  J0006$-$0015 and J1438$+$0341 have
intervening candidate
DLA systems; the latter at a redshift which places \ion{Al}{3} absorption atop
the \ion{C}{4} emission line (see Figs.~\ref{fig:gtr1}--\ref{fig:gtr2}).
J1245$-$0027 has an associated \ion{C}{4} absorber which may be intrinsic, as
the absorption appears saturated even though the flux does not decrease to zero.
J1737$+$5828 may have a weak associated \ion{C}{4} absorber.
None of these narrow intervening or associated absorption systems 
is likely to have column densities of gas or dust large enough
to affect significantly the colors or X-ray properties of our targets.

\subsubsection{J0156$+$0053 ($z=1.652$)}
The ultraviolet spectrum of this quasar shows unusually strong, narrow
\ion{He}{2} and \ion{O}{3}] emission lines as well as the reddest
  \delg\ and the next-to-smallest blueshift in the sample.  
  In addition, it has the smallest inferred $\Gamma$
  (Tables~\ref{tab:log} and \ref{tab:xcalc}).  This quasar clearly
  has extreme properties in both the ultraviolet and X-ray regimes.

\subsubsection{J0200$-$0845 ($z=1.942$)}
\label{sec:bal}
We initially identified archival quasar observations based only on the
overlap between the SDSS DR1 quasar catalog and the \chandra\ archive.
Subsequent checking revealed J0200$-$0845 to be a radio-loud 
BAL quasar with 
radio-loudness parameter, $R_i=1.66$ \citep{ivezic+02}.
Given the known connection between
both radio-loudness \citep[e.g.,][]{ReTu2000} and the presence of broad ultraviolet absorption 
lines \citep[e.g.,][]{gall+99} to broad-band X-ray properties, this object is not appropriate for our
study, and has been excluded from all correlation analysis and X-ray
spectral fitting.  Furthermore, intrinsic UV absorption may make the
blueshift measured by the SDSS pipeline inaccurate, and the \delg\
could be affected by reddening. We have included this object in the data
tables to present the X-ray properties for the record.

\subsubsection{J1151$+$0038 (LBQS1148$+$0055; $z=1.884$)}
Visual inspection of the \chandra\ data revealed two X-ray point
sources within 4\arcsec\ of the SDSS optical position of this
quasar.  The fainter X-ray source is coincident with the optical
position of a lower-luminosity quasar, LBQS1148$+$0055B, with a
discordant redshift \citep[$z=1.409$;][]{hew+98}.  This source is
spatially resolved from J1151$+$0038 and does not contaminate
the X-ray analysis.  LBQS1148$+$0055B is not included separately
in our study because it has no SDSS spectrum.

\section{Trends with Blueshift and Observed-Frame Color}
\label{sec:trends}

It has been shown that \aox\ is correlated with UV luminosity
(e.g., \citealt{Green1995,Vignali2003} and references therein). 
We choose to remove that correlation
and look for trends as a function of \daox=\aox$-$\aox$(L_{\rm
  2500})$, where \aox$(L_{\rm 2500})$ is the expected \aox\ for the
observed luminosity, $L_{\rm 2500}$, of the quasar, taken from 
Equation 4 of \citet{Vignali2003}.  The value of \daox\ reveals
an excess or deficit of X-ray emission relative to the UV luminosity;
negative values of \daox\ indicate a deficit of X-ray emission.
For reference, the range of $L_{\rm 2500}$ in our sample is approximately 1 dex
(Table~\ref{tab:xcalc}).   
We find no correlation of \daox\ with \GHR,
with \ion{C}{4} blueshift (Figure \ref{fig:pat1}a),
or with \delg\ color (Figure~\ref{fig:pat1}b).  
We use \GHR\ rather than \HR\ to remove any systematic effects
introduced by differences of the ACIS response as a function of
position on the detector or observation date. We use \delg\ rather
than simply $g-i$ because the observed $g-i$ color of a typical quasar
varies with redshift.  The parameter \delg\ is the observed
$g-i$ minus the median of the DR1 $g-i$ distribution at the redshift of the quasar \citep{rfs+01}.
The three line segments in Figure\,\ref{fig:pat1}b 
show how absorption and reddening by neutral gas and dust 
with column densities \nh=7.8$\times$10$^{20}$\cmsq, 
2$\times$10$^{21}$\cmsq, and 3.5$\times$10$^{21}$\cmsq\ 
at $z \simeq 2$ would affect \aox\ and \delg,
	using the SMC dust-to-gas ratio of \cite{bou+85}.  
\citet{ric+03} have argued that quasars have an {\em intrinsic}
distribution of $-0.3 < $\delg$ < 0.3$ and that quasars with
\delg$\gtrsim 0.3$ are dominated by objects reddened by SMC-like
dust \citep[see also][]{hopkins04} rather than by intrinsically red objects. 
As seen in Figure~\ref{fig:pat1}b, gas with SMC-like dust would have a
much larger effect on the \delg\ color than on \daox.  

To investigate potential statistically significant trends, we
calculated both Spearman's $\rho$ and Kendall's $\tau$, two
non-parametric measures of the probability of significant correlations
for \daox\ and \GHR\ versus blueshift (Figs. \ref{fig:pat1}a and c) and
\daox\ and \GHR\ versus \delg.  The only marginally significant trend we detect is an inverse correlation of \GHR\
with \delg\ whereby redder objects
tend to have harder \mbox{0.5--8.0~keV} X-ray spectra. The probability
that \GHR\  is uncorrelated with \delg\ is 0.039 and 0.028 for the
Spearman and Kendall tests, respectively.  

\citet{ric+02} determined that the range of \delg\ is dominated by 
differences in the intrinsic 
optical/UV continuum spectral index; this spectral index is a more precise
measure of the continuum color than \delg.  
We fit the SDSS spectra between 1450 and
2200\,\AA\ with a power law model, $f_{\nu}\propto \nu^{\alpha_{\rm uv}}$,
to determine the spectral index, \alphauv\ (see Table~\ref{tab:opt}).
We then tested \GHR\ versus \alphauv.  These two parameters are
more significantly correlated than \GHR\ and \delg;  the
probability of no correlation is 0.002/0.003
for the Spearman/Kendall tests, respectively.

For a sample of narrow-line Seyfert 1s (NLS1s), \citet{lm04} found a
suggestion that large-blueshift objects tended to have softer X-ray
spectra.  By eye, there is also a possible correlation of \GHR\ with
blueshift for our sample.  However, at present the data do not support a claim
of a significant correlation; the Spearman/Kendall probabilities that
these parameters are uncorrelated are 0.33/0.33. 
Furthermore, quasars with the highest blueshifts have a bluer color
distribution than those with smaller blueshifts \citep{ric+02}, and so
optical/UV color may be the primary driver for the observed trend.

\citet{lm04} also find a weak anticorrelation between \aox\
and the fraction of the \CIV\ line blueward of the rest-frame center
(another measure of \CIV\ blueshift) for the NLS1s.  Our data
(Figure~{\ref{fig:pat1}a}) do not support such a correlation for
quasars.

Throughout this paper, for those correlations which have Spearman and
Kendall probabilities of no correlation $<0.03$, we also perform a
linear regression using the Bivariate Correlated Errors and intrinsic
Scatter (BCES) estimator method \citep{ab96}.  The BCES estimator
takes into account errors in both parameters as well as intrinsic
scatter in calculating the best-fitting slope, intercept, and
1$\sigma$ errors in each.  The best-fitting line for the \GHR--\delg\
correlation is plotted in Figure~\ref{fig:pat1}d as well as lines
bracketing the 1$\sigma$ errors in the slope and intercept.

\section{Joint X-ray Spectral Fitting}
\label{sec:joint}

Given the uncertain interpretation of \GHR, which is a coarse spectral
parameterization, we proceed to joint-spectral fitting.
This technique enables utilization of all of the spectral information available in these
exploratory \chandra\ observations to determine the average properties
of a sample. The model fit in each case is a power-law continuum with both
Galactic and intrinsic neutral absorption. 
For each quasar, the Galactic \nh\ and $z$ are fixed to
the appropriate values, the values for neutral, intrinsic \nh\ and $\Gamma$ are tied
to the other quasars in the sample, and the values of $N_{\rm 1keV}$ are
free to vary for each quasar.  

Each spectrum was extracted from the source cell used for aperture
photometry, and an arf
and rmf were generated using the CIAO 3.0 script {\em psextract} which appropriately modifies
the arf to take into account the low-energy quantum efficiency
degradation. Because of the low number of counts in each spectrum, the
data were fit by minimizing the $C$-statistic \citep{Cash}, an option
for low-count spectra within {\em XSPEC}.  For this type of fitting,
the data are not binned and background spectra are not subtracted in
order to maintain the Poisson nature of the data.  Errors given in the
text are for $90\%$ confidence for two parameters of interest, $\Delta
C=4.61$, unless otherwise indicated. 

\subsection{Average X-ray Spectral Properties of Blueshift Samples}

We first fit the new data for the large-blueshift
($\Delta v_{\rm b}>1100$\kms) and small-blueshift ($\Delta v_{\rm
  b}<500$\kms) quasar samples separately.   
This approach yields the average spectral properties of these
representatives of the extreme ends of the blueshift distribution.
The total 0.5--8.0~keV counts in the large- and small-blueshift
samples were 129 and 180, respectively.

For the large-blueshift group (sample 1), the best-fitting photon index,
$\Gamma=2.0\pm0.6$, is consistent with the average of a sample of typical radio-quiet quasars
\citep[$2.0\pm0.3$; e.g.,][]{GeoEtal2000}, while for the small-blueshift
group (sample 2), the best-fitting photon index was noticeably flatter:
$\Gamma=1.4\pm0.3$. We consider this latter result uncertain as
J0156$+$0053, with 89 counts and \GHR=$1.1\pm0.2$, 
contributes half of the signal to the joint-spectral fitting for the
small-blueshift group, and appears to drive the low value for
$\Gamma$. Nevertheless, a flatter $\Gamma$ for the small-blueshift
sample is consistent with the measured values for \GHR.
While the small-blueshift quasars have best-fitting intrinsic column densities consistent with
zero, the large-blueshift quasar spectra indicate that intrinsic absorption
is present with \nh=$(1.6^{+1.9}_{-1.5})\times10^{22}$\cmsq.  
With the sample data quality and the observed-frame
0.5--8.0~keV ACIS bandpass, our joint-spectral fitting is not
sensitive to the ionization state of the gas, and ionized gas would have
higher column densities. 

The binned spectra presented in Figure~\ref{fig:xspect} illustrate 
the difference between a flat power-law continuum and a
steeper power-law continuum with absorption.  Two composite spectra
were created, one each of the large-blueshift and small-blueshift
samples, by adding together the unbinned spectra in each sample.
Three arfs were combined by 
weighting by the \mbox{0.5--8.0} keV counts in each spectrum, to create a
composite arf.  The same weighting scheme was used to make composite
rmfs.  The data were grouped to have at least 5 counts per bin.
The data below 1~keV were ignored, and both spectra were fit
independently with power-law models.  This procedure sets both $N_{\rm 1keV}$
and $\Gamma$ of the hard-band continuum and is insensitive to
\nh$\lesssim3\times10^{22}$\cmsq\ at these redshifts. 
Finally, the power-law models were extrapolated to 0.5 keV
for comparision with the 0.5--1 keV data.
As can be seen in the residual panel in Figure~\ref{fig:xspect}, while
a $\Gamma\sim1.4$ power-law
model fits the soft-band small-blueshift composite spectrum quite well, a
$\Gamma\sim2.2$ power-law model significantly overpredicts the
soft-band counts from the large-blueshift composite spectrum.  
These negative residuals are a clear signature of intrinsic absorption.
For reference, uncertainties in the contaminant model applied to
correct the arfs for the low-energy quantum efficiency degradation are
estimated to be $\lesssim5\%$ in the 0.5--1.0 keV
range,\footnote{http://asc.harvard.edu/cal/Acis/Cal\_prods/qeDeg/index.html}
much less than these large negative residuals.  Furthermore, any errors
in the contaminant model would not affect the large-blueshift sample
preferentially, as the observations contributing to each sample span roughly the same time.

To confirm these trends, we included the archival spectra in the
joint-spectral fitting, and created a third, `moderate' blueshift group with
$\Delta v_{\rm b}=500$--1100\kms.  We excluded the archival quasar
J1245$-$0027 because of the large uncertainty 
in its blueshift 
(see Table~\ref{tab:opt}). The same procedure was
followed, with the large (sample 3), small (sample 4) and moderate
(sample 5) blueshift groups.  The addition of the archival quasars
significantly increased the total 0.5--8.0~keV counts in the large- and small-blueshift
samples to 266 and 260, respectively.  Once again, best-fitting
parameters for the large-blueshift sample included a steeper photon index,
$\Gamma=2.0^{+0.8}_{-0.3}$, and significant intrinsic absorption,
\nh=$(1.5^{+2.3}_{-0.9})\times10^{22}$\cmsq.  The small blueshift sample
also showed consistent results, with a harder X-ray continuum,
$\Gamma=1.6^{+0.3}_{-0.2}$.  At $90\%$ confidence, intrinsic
absorption is constrained to be $<0.63\times10^{22}$\cmsq. The
upper limit is set primarily by the low-energy cutoff (0.5~keV
observed-frame) of the ACIS bandpass.

The contours for samples \mbox{3--5} are
presented in Figure~\ref{fig:contours}, and the results from the
joint-spectral fitting of samples 1--5 are listed in
Table~\ref{tab:spec}. The moderate-blueshift group, sample 5, has only three quasars, but with
the total number counts equal to 1139, has significantly higher signal-to-noise
ratio than the other two samples.  While the best-fitting photon index
for this sample, $\Gamma=2.0^{+0.3}_{-0.1}$, is consistent with sample
3, the best-fitting intrinsic \nh\ is consistent with zero.  This
suggests that samples 3 and 5 have consistent intrinsic hard-band
X-ray continua, while the quasars in sample 4 may have harder
(flatter) X-ray continua. 
To investigate this possible trend in more depth, 
we pursued joint-spectral fitting analysis further.

\subsection{Blueshift or Color?}

In Figure~\ref{fig:C4dist}, we present both the histogram of \CIV\ blueshifts
for the $z=1.54$--2.20 SDSS Data Release 2 \citep[DR2;][]{sdss_dr2}
quasars (updated from Figure~1 in \citealt{ric+02}) and the plot of
\delg\ versus \CIV\ blueshift for these
objects. As noted by \citet{ric+02}, quasar composite spectra made by
binning in blueshift
indicate that color differences correlate with blueshift: 
large-blueshift quasars tend to have bluer \delg\ colors, while
small-blueshift quasars tend to have redder optical/UV continua.
However, the two-dimensional structure in the \delg--blueshift distribution
indicates that these properties are not simply related.  Though
large-blueshift quasars are much more likely to have blue \delg\
colors, quasars with blueshifts less than the median value of
$\sim800$\kms\ span the entire color range.

In an attempt to disentangle the connection
between X-ray spectral differences and
blueshift, we extended the joint-spectral fitting by binning the quasars
into finer blueshift bins.  The three moderate-blueshift
quasars (J0113$+$1535, J1438$+$0341, and J2348$+$0057) 
each had enough counts to be fit independently.  
The other quasars were grouped into the smallest
samples (from 2--4 objects; see Table~\ref{tab:opt}) 
that would enable reasonable constraints
to be set on $\Gamma$ and intrinsic \nh\ from joint-spectral fitting
of an absorbed power-law model. The results from this analysis are presented in
Figure~\ref{fig:gall}. 

Figures~\ref{fig:gall}a and \ref{fig:pat1}d
indicate that both large-blueshift and
blue quasars tend to have larger X-ray photon indices.  However,
only the second trend is statistically significant.  Therefore, 
\alphauv, rather than blueshift, appears to be linked to the steepness
of the hard-band X-ray continuum. Figure~\ref{fig:gall}b clearly 
supports the significant detection of intrinsic absorption in the
large-blueshift quasars.

\section{Summary of X-ray Analysis and Comparison to BQS Quasars}
\label{sec:s}
With a modest amount of \chandra\ exposure, we have derived some
insight into the connection between X-ray and UV spectral properties
of luminous quasars. From joint-spectral fitting, we have found that
the quasars
with large blueshifts ($\Delta v_{\rm b}\gtrsim1100$\kms) in our
sample show significant evidence for intrinsic absorption with
\nh$\sim10^{22}$\cmsq\ (assuming neutral gas with solar abundances). 
For this sample, unlike with samples of low
$z$ or BAL quasars, \GHR\ is not a sensitive absorption indicator.
The combination of column density
(\nh$\sim10^{22}$\cmsq) and redshift pushes the energy cutoff
from absorption to observed-frame $\sim0.8$~keV (see
Fig.~\ref{fig:xspect}). The effective
area of ACIS S3 peaks between 1--2~keV, after the spectrum has
recovered.  The low-energy curvature of the power-law spectrum from
absorption is thus only evident with the additional energy resolution
utilized in joint-spectral fitting.  These same factors, bandpass,
redshift, and column density, also make \aox\ a weak indicator of intrinsic
absorption for this sample, unlike for the \rosat\ survey of low-redshift
BQS quasars by \citet{blw00}.
For that sample, \aox\,$\lesssim-2.0$ was a strong predictor of
comparable column densities of X-ray absorption \citep{gall+01}.

Unfortunately, the BQS sample cannot be used effectively to study
X-ray absorption in large-blueshift quasars.
Of luminous ($M_V < -24$), radio-quiet BQS quasars with measured blueshifts \citep{bl04b},
only two would qualify for our large-blueshift sample with $\Delta
v_{\rm b} \gtrsim 1100$\kms.  The first, PG~1259$+$593 ($\Delta v_{\rm
  b}= 3304$\kms), was not detected with
\rosat, and \citet{blw00} measured an upper limit on \aox\ of $-1.79$,
which is (at least) moderately X-ray weak.  For the second,
PG~1543$+$489 ($\Delta v_{\rm b}= 2032$\kms),
\citet{GeoEtal2000} find their preferred spectral model for the \asca\ data 
to include intrinsic absorption (\nh$=0.7\times10^{21}$~\cmsq\ for
neutral gas) which may be ionized.  Though the X-ray properties of
these two quasars suggest that both may harbor intrinsic absorbers,
the quality of the existing X-ray data is not sufficient for such a claim.

Our second result is that \GHR\ correlates 
with \alphauv.
This might be expected if smaller values \GHR\ were tracing intrinsic
absorption. Instead, the results from joint-spectral fitting indicate
that the \GHR--\alphauv\ trend seen in
Figure~\ref{fig:pat1}d  is {\em not} driven by absorption: instead the bluest
quasars with the largest \GHR\ values show evidence for
absorption. The correlation indicates an inherent difference in the
actual X-ray photon index as a function of color, with redder quasars having
harder X-ray spectra.

Given that our sample is small and the statistical errors in the X-ray
spectral properties are large, we investigated the connection between
optical/UV continuum color and $\Gamma$ further with additional data from \citet[][hereafter P04]{porquet04}.
P04 systematically analyzed BQS X-ray spectra
from the \xmm\ archive.  Many of these quasars also have measurements
of their optical/UV continuum slopes, $\alpha_{\rm o}$, from power-law fits to
narrow-band optical photometry by \citet{neug87}. 
For comparison with our results, the P04
sample was stripped of all radio-loud quasars and all quasars more than 2 magnitudes 
less luminous than PG 0953$+$396, the most luminous radio-quiet quasar in 
the sample ($M_V=-25.65$; Table 1 of P04), because both radio-loudness and
luminosity might influence X-ray properties.  The resulting range
of optical luminosity matches that
of our sample, though our sample is significantly more luminous
overall (see column 5 of Table~{\ref{tab:opt}). Of the 16 remaining quasars, 14 had
 measurements of $\alpha_{\rm o}$. 

In Figure~\ref{fig:pg}, we plot the observed-frame 2--5~keV
$\Gamma$ measured by P04 versus $\alpha_{\rm o}$ from
\citet{neug87}.  Many of these quasars have additional hard-band $\Gamma$
measurements from spectral fits to different data (see the Appendix of
P04 and reference therein), and we plot these
as well with filled circles.  Several quasars (PG~1202$+$281, 1307$+$085, 1353$+$183,
and 1613$+$658) show significant differences in $\Gamma$ between
observations, which may result from actual variability and/or
differences in analyses or observatories. There is clearly a large scatter, but the general
trend that the bluest quasars have the steepest hard-band X-ray
spectra is consistent with our results. For the filtered P04 sample,
the Spearman and Kendall
probabilities of no correlation between $\Gamma$ and $\alpha_{\rm
o}$ are 0.038 and 0.028, respectively.  If the averages between the
P04 and previous values are used, the no-correlation probabilities are 0.022 and
0.014 for the same sample.  The BCES estimator linear fit to the
average $\Gamma$--$\alpha_{\rm o}$ datapoints is overplotted in Figure~\ref{fig:pg}.
To compare directly the BQS data to the data plotted in
Figure~\ref{fig:pat1}d, 
a polygon encompassing the 1$\sigma$ range in slope and intercept for
the linear fit is overplotted in
Figure~\ref{fig:pg}. The two independent fits are consistent within
the 1$\sigma$ errors in both slope and intercept.

Neither of these results relating hard-band $\Gamma$ to optical/UV
color is independently conclusive, but the
combined evidence from the \GHR--\alphauv\ trend
and the presented BQS data points to a consistent picture.  The
hard-band X-ray and optical/UV continua are linked; the bluer
quasars exhibit steeper X-ray spectra.

\section{Discussion}
\label{sec:d}

The observation that the large-blueshift quasars are blue in the
optical/UV (Figure~\ref{fig:C4dist}b) and have higher intrinsic
absorption column densities (Figures~\ref{fig:contours} and
\ref{fig:gall}) is consistent with large blueshifts occurring in
quasars observed close to the plane of the accretion disk. In
particular, the models of \citet{hab+00} predict (their Fig.~12) that
edge-on accretion disks are bluer than face-on disks.
Such large inclination angles with respect to the disk normal would be
expected to yield larger X-ray absorption column densities if the
absorbing material is found closest to the disk.  
This scenario would also be consistent with the idea that BAL~quasar outflows
  (known to have significant absorption in both the UV and X-ray)
are equatorial. In addition, \citet{rrh+03} suggest that BAL quasars
are also {\em intrinsically} blue.

If we extend this connection, large-blueshift quasars
may have orientation angles close to those of BAL~quasars that do not actually
intercept the UV-absorbing wind.  The absence of significant \CIV\
absorption in the large-blueshift quasars
indicates that the X-ray absorbing gas is either highly ionized (with
little UV opacity) and/or does not obscure the UV continuum.  The
fraction of carbon ionized less than \ion{C}{5} drops to $<1\%$
in a photoionized plasma with $\xi\gtrsim8$\,
erg\,cm\persec (where
  $\xi=L/nR^{2}$ [$L$ is the integrated luminosity from 1 to 1000
  ryd, $n$ is the gas density, and $R$ is the distance from the
  radiation source]; \citealt{kb01}).  At this ionization parameter
for $z\sim2$, the
actual column density of solar metallicity gas giving the same opacity in the observed \chandra\
bandpass as 1.5$\times10^{22}$\cmsq\ of neutral gas is $\sim50\%$ larger.  

The X-ray absorption
seen in the large-blueshift quasars could be identified with the shielding
gas postulated by \citet{mcgv95}.  This gas is required in their 
model to prevent soft X-rays from over-ionizing the disk-wind gas.
Without it, radiation pressure by UV resonance line photons cannot
radiatively drive the wind to the high velocities seen in BAL
quasars.  If this interpretation is correct, these data suggest that
the shielding gas may have a larger covering fraction than the BAL
outflow, and the blueshifted \CIV\ emission could be interpreted as a disk-wind
signature.  
For reference, BAL quasars typically
show power-law X-ray spectra ($\Gamma\sim2$) with complex, intrinsic
X-ray absorption of \nh\,$\gtrsim10^{23}$\cmsq\
\citep{GreenEtal2001,GallagherEtal2002}.

While the connection between extreme blueshift and X-ray absorption fits
reasonably within the \citet{mcgv95} disk-wind paradigm as an effect
of orientation, understanding the relation between \alphauv\ and $\Gamma$
is more complicated.  It is important to distinguish between this survey
and previous work with \rosat\ that focused on the observed 0.1--2.0~keV
spectral slope, typically of $z\lesssim$1 quasars.  Correlations found
with \rosat\
\citep[e.g.,][]{puch+96} are much more sensitive to small absorbing
column densities, and could also be driven by the soft excess.  The
soft excess, undetectable with these data, can often be modeled as a thermal
black body, and is believed to originate in the inner accretion disk.  
For more relevant comparisons, there are many claims in the literature
of correlations with the 2--10~keV $\Gamma$
\citep[e.g.,][]{zdz99,ReTu2000,dai+04,wwm04}.
The claim of a correlation between $\Gamma$ and $L/L_{\rm Edd}$
of \citet{wwm04} is the most intriguing in the
context of our sample. 

Based on comparisons with black-hole binaries
\citep[e.g.,][]{zg04}, one might expect that objects with high
accretion rates relative to Eddington will have softer hard X-ray
spectra and a UV component that is more dominated by the disk (and
thus bluer).  {In contrast,} 
lower accretion-rate objects will have harder X-ray spectra and may appear
redder due to a weaker inner disk component to the big blue bump.
This picture is grossly consistent with the correlation of \alphauv\
with $\Gamma$.

Our two most significant
results (the presence of intrinsic X-ray absorption at large blueshift
and harder X-ray spectra correlating with redder colors) 
thus are consistent with
both orientation {\em and}
accretion rate effects (either independently or together).  
The two dimensional structure in the
\delg--blueshift distribution 
(Fig.~\ref{fig:C4dist}b) implies that this is (at least) a two
parameter problem.  

The absorption seen in the X-ray spectra of the largest-blueshift objects
suggests that we are looking ``down the wind'' in such objects; in other
words, that they are observed at the most extreme inclination angles
with respect to the disk normal possible for an optically thick wind
before BAL signatures are manifested. 
Large-blueshift objects also tend to have weak \CIV\ equivalent widths.
Therefore, the association of 1) high accretion rates and small \CIV\ 
equivalent widths \citep{bl04}, 2) BAL quasars with bluer intrinsic spectra and
large blueshifts \citep{rrh+03}, and 3) BAL quasars with high accretion rates
\citep{bor02,yw03} means that the extreme population that shows bluer
UV continua, large blueshift, weaker \CIV, and absorbed X-ray spectra may
represent the most extreme accretors with the largest inclination angles
possible for a given wind geometry.  Such a scenario would be qualitatively
consistent with the narrowing of the color/blueshift distribution toward
large blueshift velocities seen in Figure~\ref{fig:C4dist}b.

However, moving to less extreme blueshifts, the situation in this
scenario gets complicated, because both objects with less powerful winds (because
they are less luminous though still active accretors or because they
are accreting less actively) at extreme angles and objects with very
powerful winds at less extreme angles contribute.  This leads to the
stretch in the \delg\ color distribution (the ordinate in
Figure~\ref{fig:C4dist}b), which is widest close
to $\sim0$\kms\ blueshift.  Here we assume that the intrinsic \delg\
color is a marker of Eddington accretion rate, with bluer continua resulting from
higher accretion rates at a given orientation angle. 
This speculative interpretation of the two dimensional \delg--blueshift
distribution can be tested with observing programs tuned to isolate
the two phenomenological parameters, color and blueshift, in an attempt to
map them onto the underlying physical drivers.  
In this interpretation, we conjecture that choosing a narrow range of color and
luminosity would allow blueshift to be an orientation indicator,
whereas choosing a range in blueshift and luminosity allows color to
be an indicator of accretion rate (because this would remove any
orientation dependence of the color).  

Alternately, \citet{lei04} suggests that blueshifted emission comes
from a wind which arises in AGN only under certain conditions.
If blueshifts always indicate the presence of winds, the histogram in
Figure~\ref{fig:C4dist}a implies that nearly all broad-line quasars
have winds, and our detection of X-ray absorption in large-blueshift
quasars is broadly consistent with this model.  
However, simply invoking a wind
from the near face of an optically thick accretion disk to explain 
large \CIV\ blueshifts
(see also \citealt{elv04}) does not necessarily yield a simple shift,
nor does it trivially explain the weakness of the lines found in
large-blueshift quasars.  

Lastly, the relationship between the blueshifts and the \CIV\
equivalent widths (EWs) deserves further discussion.  
Any successful model of the broad line region must account for both
the blueshifts and the EWs of the emission lines.  For example, it may be
reasonable to unify the \citet{bal77} effect and emission line
blueshifts, especially given the findings by \citet{ric+02} that the
\CIV\ blueshift effect is not apparently dominated by luminosity,
by \citet{bl04} that the Baldwin effect is driven by $L/L_{Edd}$
rather than $L$, and by \citet{fk95} that the Baldwin effect is
strongest in the red wing of the \CIV\ emission line.  Perhaps as $L/L_{Edd}$
increases, the greater dominance of radiative driving over gravity produces a
faster wind with a larger opening angle relative to the disk normal.  Such
winds will cover less of the sky from the point of view of the continuum
source, thereby intercepting less ionizing radiation and lowering the EW of
\CIV\ and other lines emitted by the wind.  

\section{Conclusions}
\label{sec:conclusions}
From our analysis of \chandra\ ACIS observations of 13 radio-quiet SDSS quasars
and their optical/UV spectral properties, we present the following conclusions:

\begin{itemize}
\item{Those quasars in our sample with \CIV\ blueshifts
  $\gtrsim1100$\kms\ show evidence from joint-fitting of the X-ray
  spectra for intrinsic X-ray absorption with \nh\,$\sim10^{22}$\cmsq.  
 We interpret the presence of X-ray
  absorption in the large-blueshift sample as support for the
  orientation interpretation of the \CIV\ blueshift put forth by
  \citet{ric+02} whereby large-blueshift quasars are seen at
  inclination angles close to the line of sight through the wind
  (which may have a range of opening angles).  This result is broadly consistent with
  the disk-wind model of quasar broad-line regions of
  \citet{mcgv95}.}

\item{We find that there is a trend of steeper hard-band X-ray
  continua with bluer \alphauv\ spectral index in our sample; this result is
  supported by a complementary analysis of independent Bright Quasar Survey data from
  the literature.  We find the Eddington accretion rate, $L/L_{\rm Edd}$, to be a
  likely candidate for the primary physical driver of this trend.
}

\end{itemize}

Extending this study to larger samples with higher-quality X-ray data
is certainly warranted to test these claims. Specifically, given the
lack of \CIV\ absorption in the SDSS spectra of the large-blueshift
sample, we predict that high quality X-ray spectroscopy will reveal
this gas to be ionized with $\xi>8$\,erg\,cm\persec.
With the large effective area and 0.3--10.0~keV bandpass, sensitive
\xmm\ observations could significantly constrain both the
column density and ionization state of the absorption for each quasar
in the large-blueshift sample.
Furthermore, high signal-to-noise ratio 
X-ray observations of a quasar sample tuned to isolate the
relationship between UV/optical and X-ray continua hold promise for
understanding the effect of Eddington 
accretion rate on quasar spectral energy distributions.

\acknowledgements
We thank Mike Nowak, Andrzej Zdziarski, and Sera Markoff for illuminating
discussions of accretion disk coronae.  We thank Ari Laor for
helpful comments and access to data prior to publication.
This work was made possible by
\chandra\ X-ray Center grant G03--4144A. Support for
SCG was provided by NASA through the {\em Spitzer} Fellowship Program,
under award 1256317.  WNB acknowledges support from NASA LTSA grant
NAG5--13035, and DPS acknowledges the support of the NSF grant AST03--07582.

\begin{figure}
\centerline{\psfig{figure=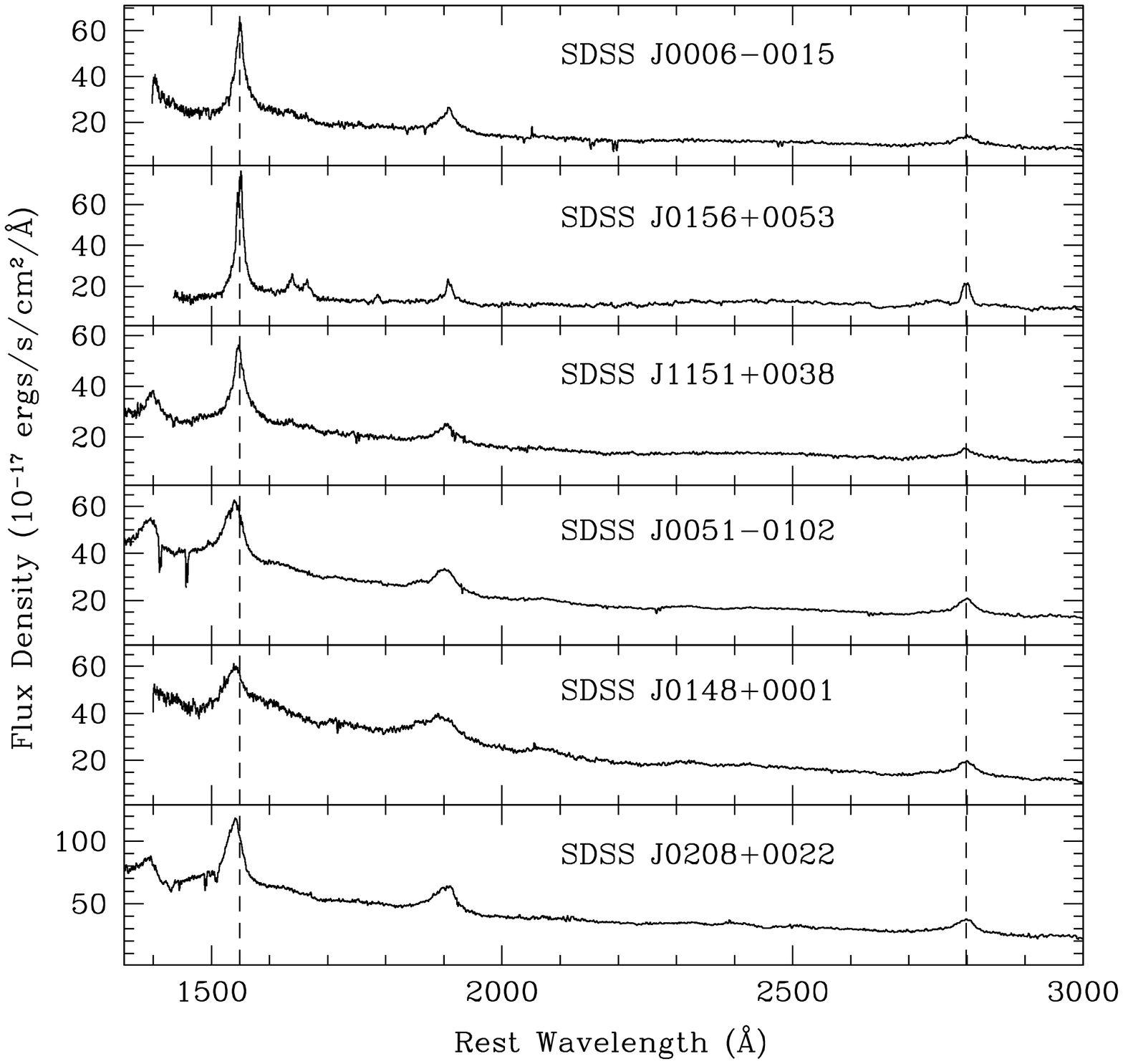,width=6.0truein,angle=0.0}}
\caption{
Spectra of SDSS quasars with new \chandra\ observations.  Dashed
vertical lines indicate the expected peaks of \CIV\ (1549.06\AA) and
\MgII\, using the \MgII\ emission line to define the systemic
redshift.
\label{fig:gtr1}
}
\end{figure} 
\begin{figure}
\centerline{\psfig{figure=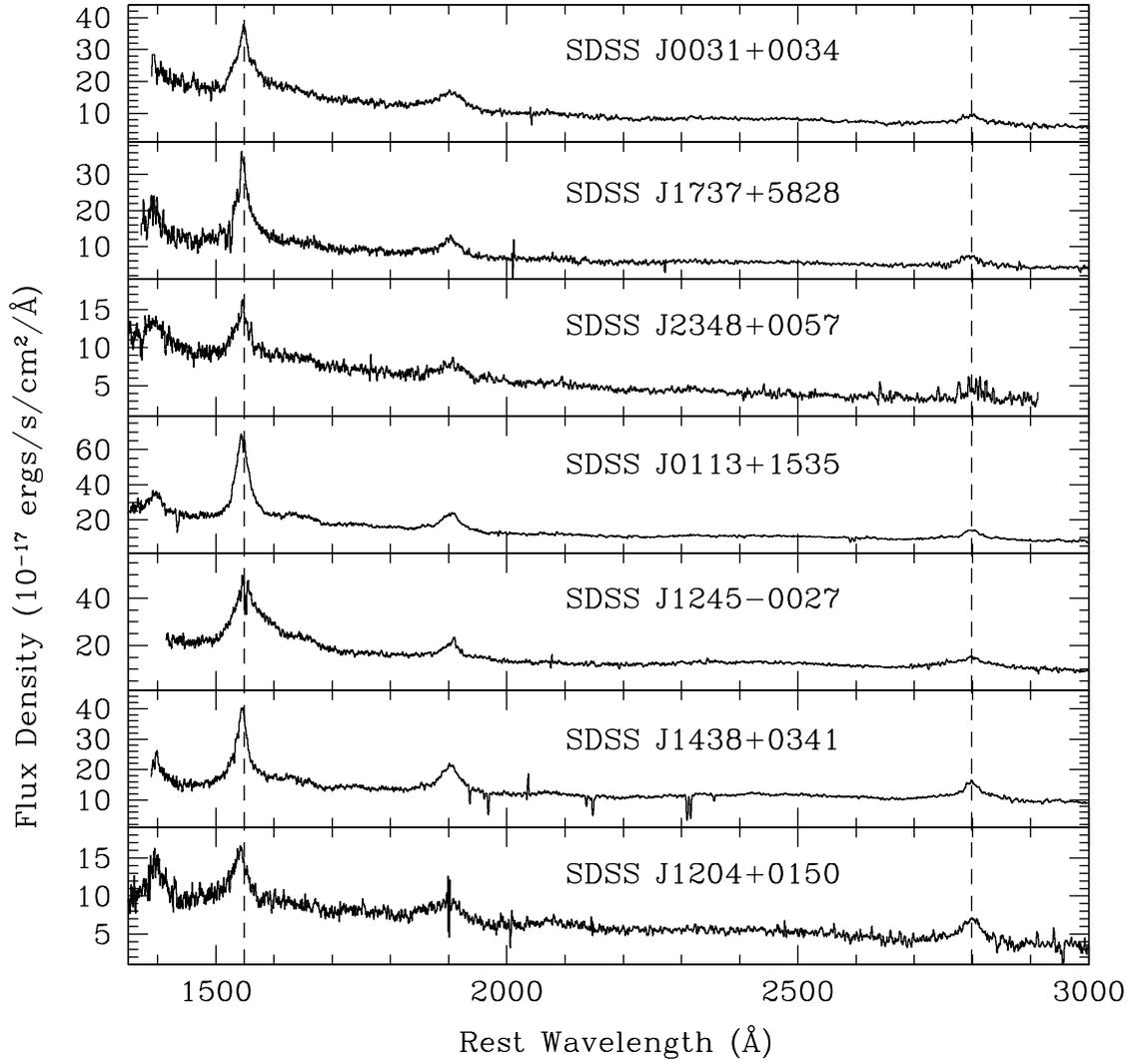,width=6.0truein,angle=0.0}}
\caption{
Spectra of SDSS quasars with archival \chandra\ observations.
\label{fig:gtr2}
}
\end{figure} 
\begin{figure}
\centerline{\psfig{figure=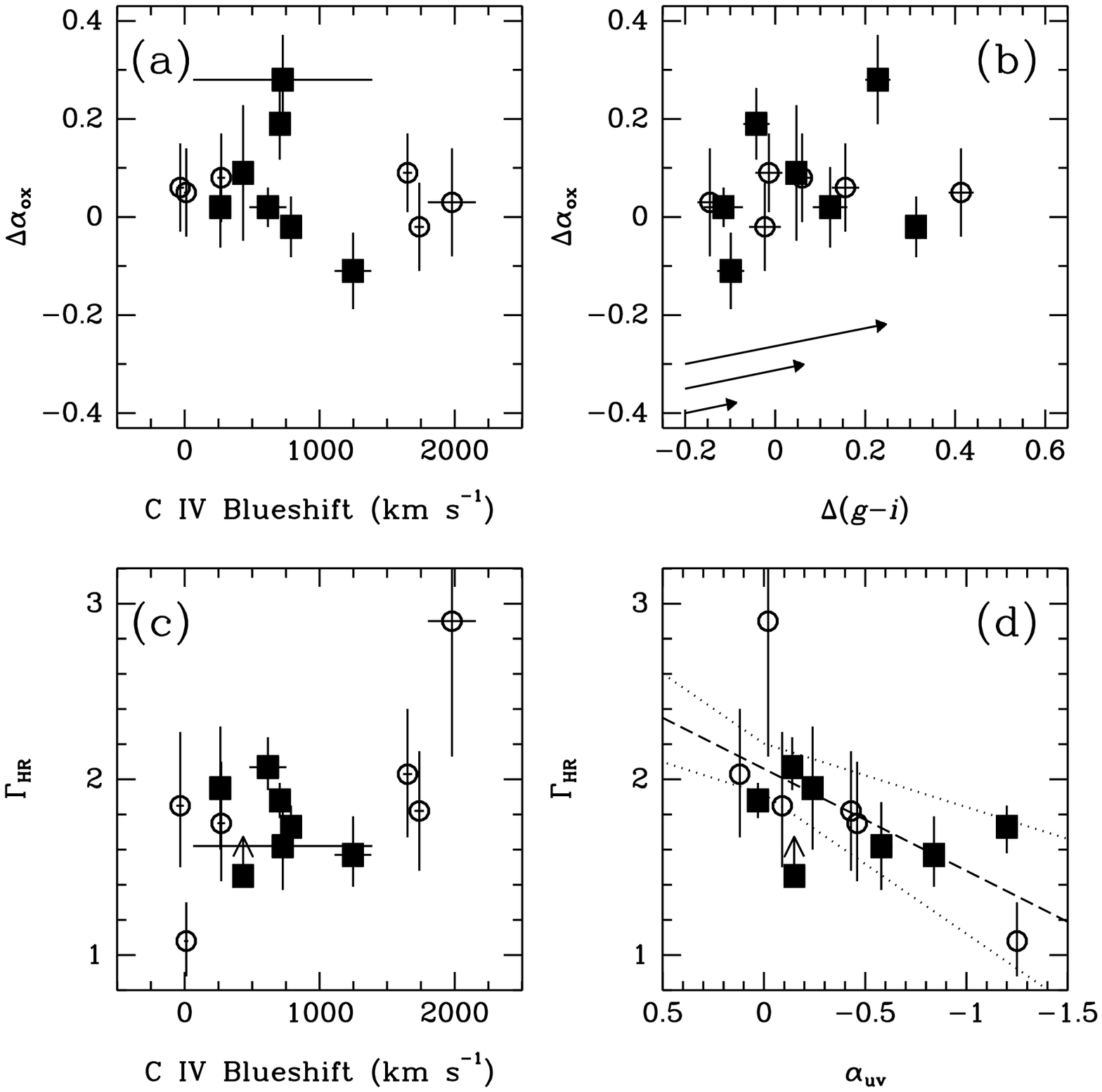,width=6.0truein,angle=0.0}}
\caption{
Observed X-ray versus optical properties of the new (open circles) and
archival (filled squares) \CIV\ blueshift sample. {\bf (a)} The value \daox\ versus \CIV\ blueshift.  The
  parameter \daox\ is the difference between the observed \aox\ and the expected \aox\
  (calculated using the luminosity density at 2500\,\AA\ from Equation 4 of \citealt{Vignali2003};
  see Table~\ref{tab:xcalc}).  The errors plotted are the errors in \aox. {\bf (b)} The
  value \daox\ versus \delg\ color. The three solid lines represent the
  expected changes in \daox\ and \delg\ from SMC reddenings of 
  $E(B-V)$=0.018, 0.046, and 0.081. 
  {\bf (c)} The photon index calculated from the hardness ratio, 
  \GHR, versus blueshift.   {\bf (d)}  \GHR\ versus \alphauv. The dashed and dotted lines indicate the
  best-fitting linear correlation and 1$\sigma$ errors in the slope
  and intercept from the BCES estimator.  
\label{fig:pat1}
}
\end{figure} 
\begin{figure}
\centerline{\psfig{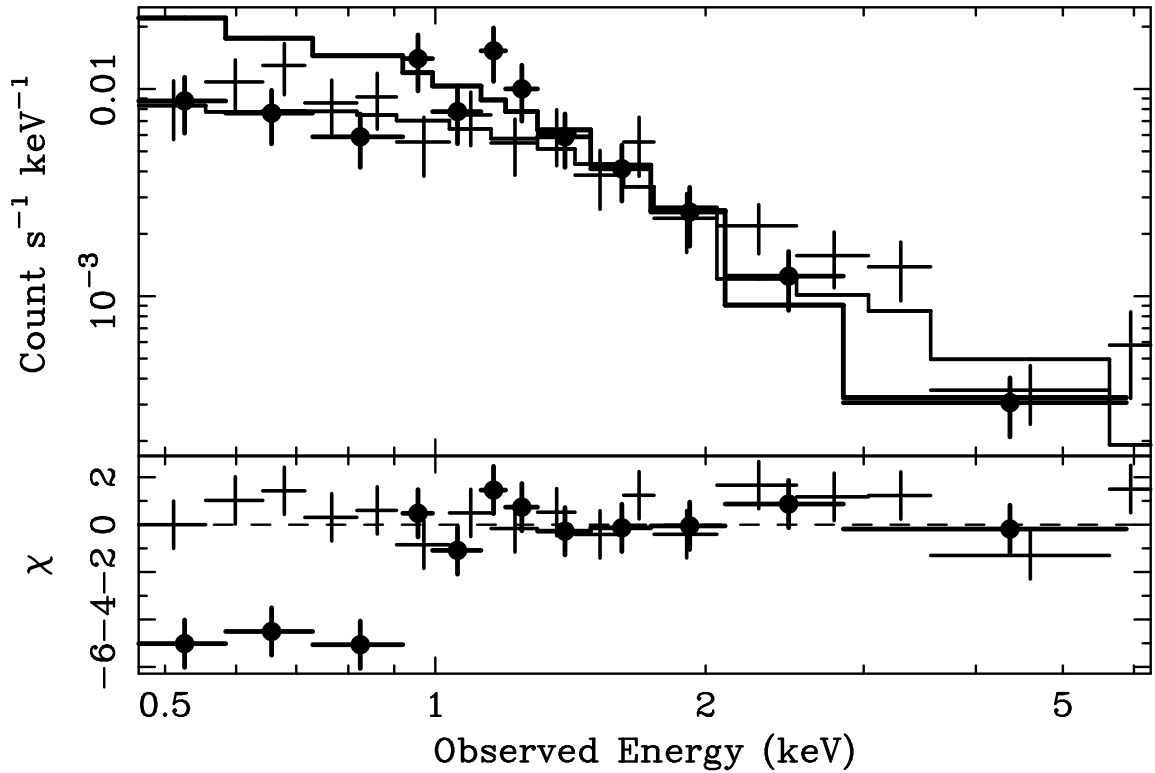}}
\caption{
\chandra\ spectra of the combined large (filled circles) and small
(crosses) blueshift samples.  Both data sets have been fit with a
power-law model above observed-frame 1~keV, and the models (solid
histograms) have been extrapolated back to 0.5~keV.  The residuals (lower panel) indicate
that while a power-law model with $\Gamma\sim1.4$ is an adequate fit
to the small blueshift sample, a power law model with $\Gamma\sim2.2$
overpredicts the counts between 0.5--1.0~keV for the large-blueshift
sample.  These negative residuals are the signature of intrinsic
absorption.
\label{fig:xspect}
}
\end{figure}
\begin{figure}
\centerline{\psfig{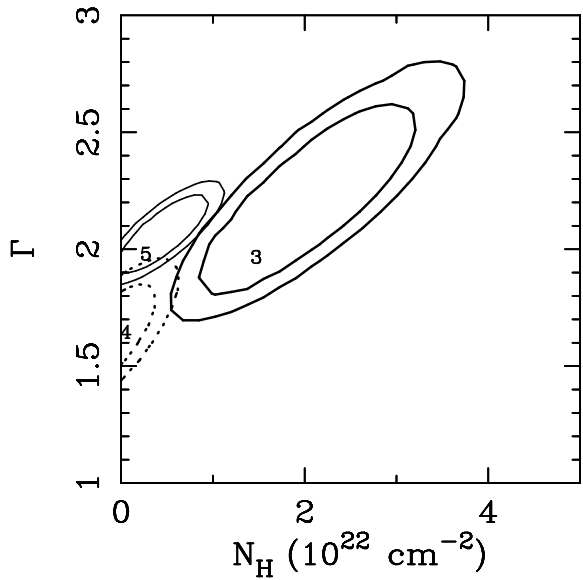}}
\caption{
Confidence contours of photon index, $\Gamma$, versus intrinsic \nh\ from
joint-spectral fitting for the new plus archival large
(thick solid curves; sample 3), moderate (thin solid curves; sample
4), and small (dotted curves; sample 5) blueshift samples as defined
in Table~\ref{tab:spec}. The contours are for 68$\%$ ($\Delta C=2.30$) 
and 90$\%$ ($\Delta C=4.61$) confidence.
\label{fig:contours}
}
\end{figure}
\begin{figure}
\centerline{\psfig{figure=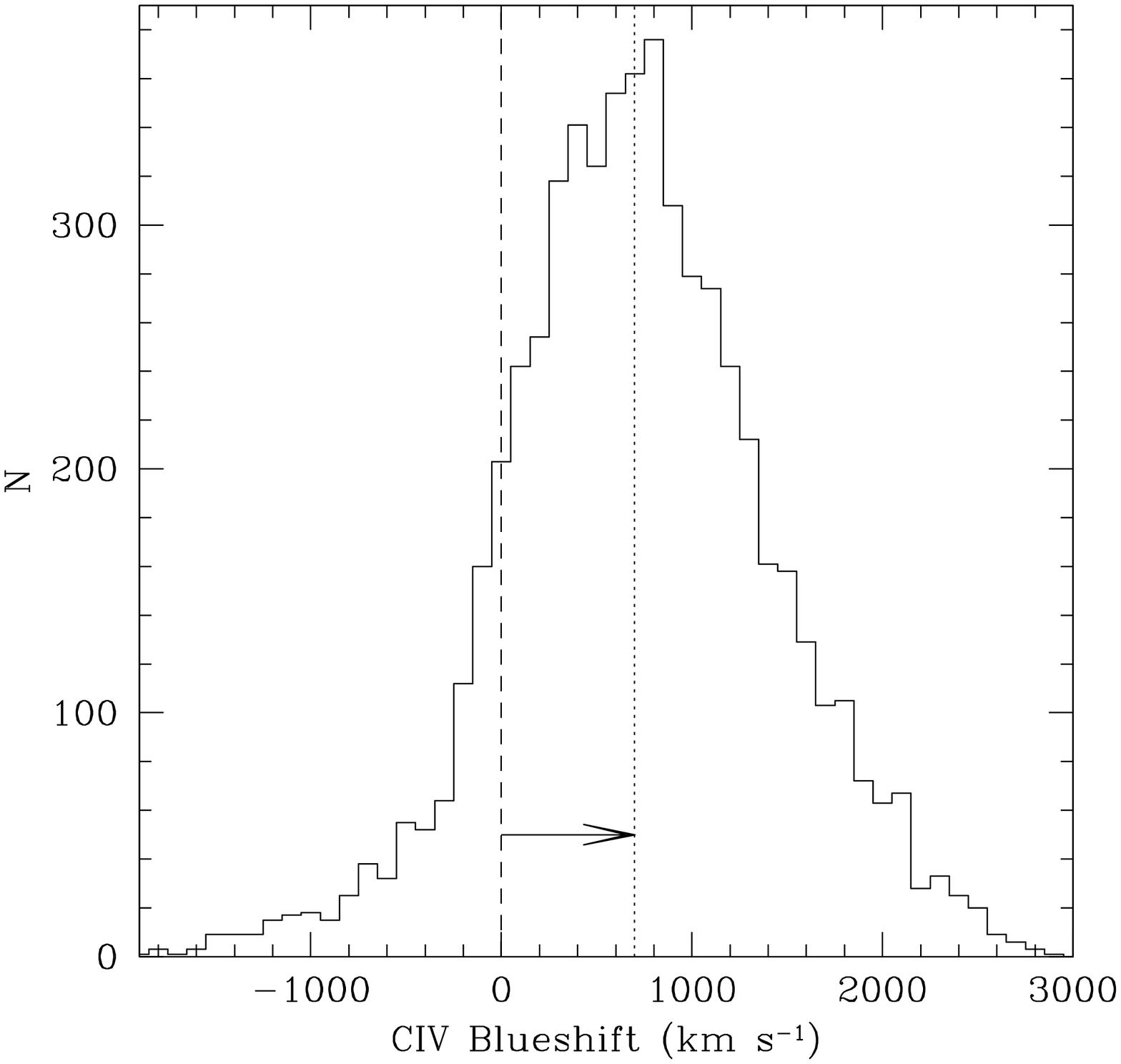,width=3.0truein}
\hfill
\psfig{figure=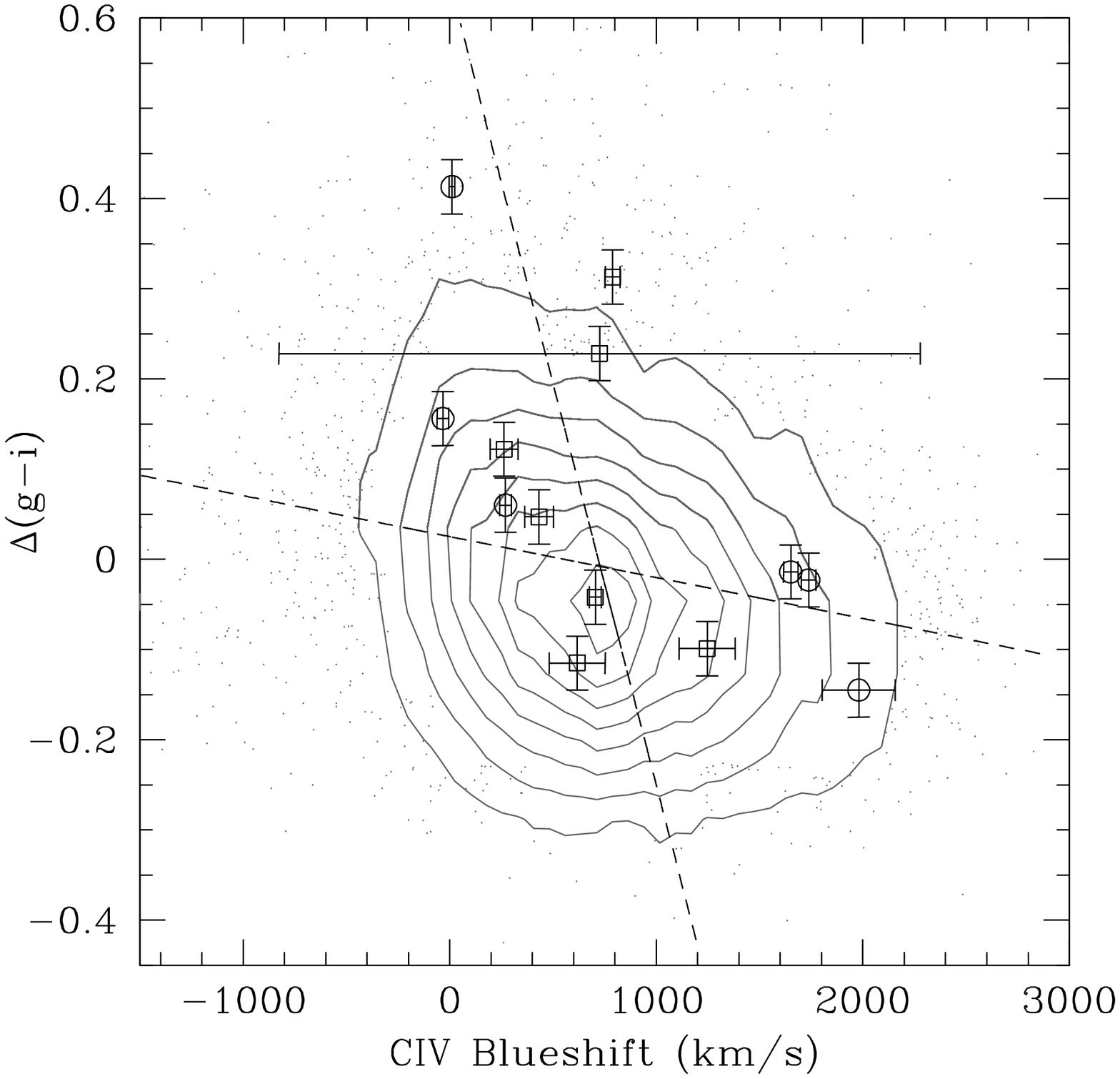,width=3.0truein,angle=0}}
\caption{{\bf Left:} Histogram of the distribution of \CIV\ blueshifts
  from the SDSS DR2 quasar sample with \CIV\ blueshift
  errors $<$250\kms.  The dotted vertical line marks the
  median blueshift value of the sample, and the dashed vertical line
  marks zero velocity.    
  {\bf Right:} Contour plot of the
  distribution of \delg\ versus \CIV\ blueshifts for the
  entire SDSS DR2 sample with $z$=1.54--2.20.  Quasars with new observations
  are indicated with open circles and archival quasars with open squares.
  The dashed lines are the linear correlations from least-squares
  fits taking either color or blueshift as the dependent variable.
  \label{fig:C4dist}
}
\end{figure}
\begin{figure}
\centerline{\psfig{figure=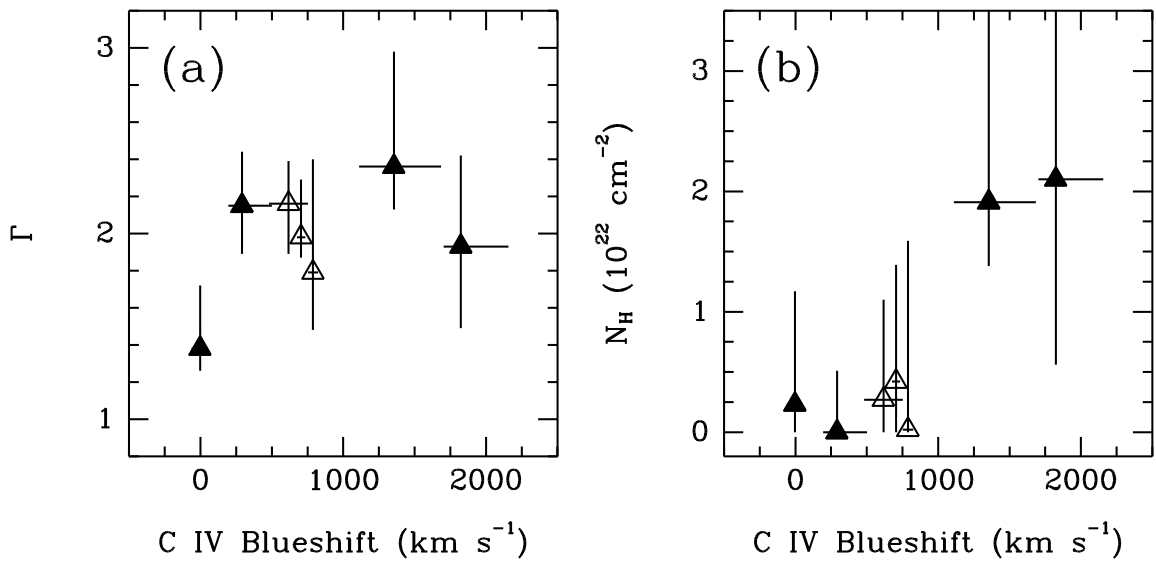,width=6.0truein}}
\caption{
  Results from X-ray joint-spectral fitting with absorbed power-law
  models: {\bf (a)} $\Gamma$ and {\bf (ab)} \nh\ versus \CIV\
  blueshift. Each datapoint
  indicates a small sample of quasars with a similar blueshift. 
  The open symbols indicate
  spectra with enough X-ray counts to be fit individually.  Errors in
  the ordinate are 1$\sigma$ taking two parameters to be of interest
  ($\Delta C = 2.3$).
  For the jointly fit quasars (solid symbols), the abscissa value
  is the average of the quasars in each sample weighted by X-ray
  counts.  Errors in the abscissa indicate the range (including
  uncertainties) of blueshift values in the objects contributing to each
  sample.  
  \label{fig:gall}
}
\end{figure}
\begin{figure}
\centerline{\psfig{figure=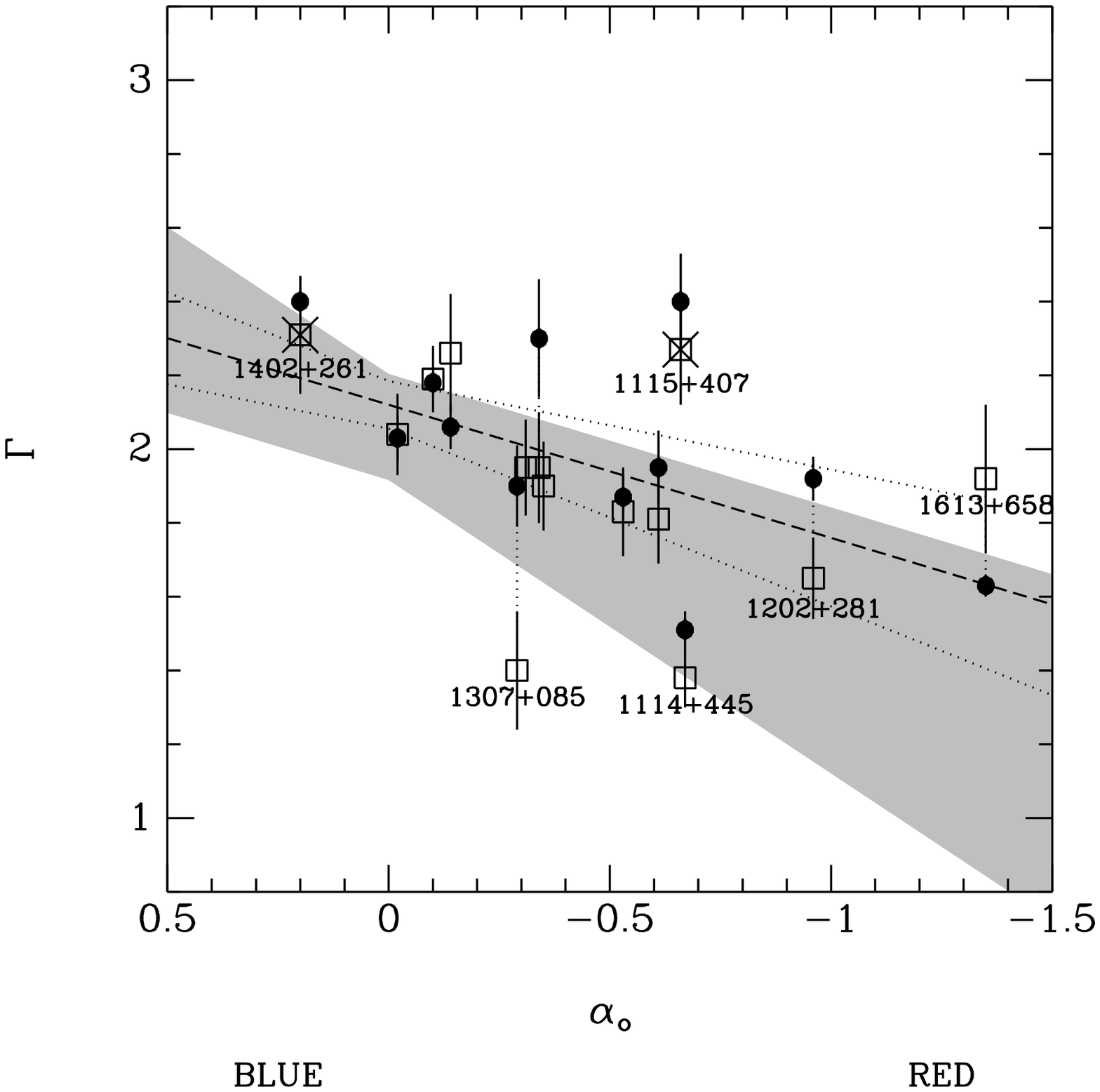,width=4.0truein}}
\caption{
Plot of $\Gamma$ versus $\alpha_{\rm o}$ for BQS quasars
in the \xmm\ sample of \citet{porquet04}. The open squares are the
\citet{porquet04} 2--5~keV $\Gamma$ values from fitting \xmm\ spectroscopic
data; the filled circles indicate other spectroscopic X-ray
measures of hard-band $\Gamma$ from the literature of some of the same quasars 
\citep[see Appendix;][]{porquet04}. Dotted vertical lines connect
datapoints for those quasars whose measures of $\Gamma$ differ by more
than the given error bars. The values of
$\alpha_{\rm o}$ were measured by \citet{neug87} from continuum
photometry.  The `$\times$' symbol indicates narrow-line quasars with
\mbox{H$\beta\, \rm FWHM<2000$\kms}.
Narrow-line quasars, quasars with large $\Gamma$ variability,
and outliers have been labeled for reference.  PG~1114$+$445 has
complex, intrinsic absorption that complicates the determination of
$\Gamma$; \citet[][]{geo97} measure $\Gamma=1.8\pm0.1$.
The dashed and dotted lines indicates the BCES estimator linear
fit to the correlation of the average $\Gamma$ to $\alpha_{\rm o}$.
The gray polygon marks the BCES estimator linear fit 
to the data for the SDSS sample plotted in Fig.~\ref{fig:pat1}d.
\label{fig:pg}
}
\end{figure}
\begin{deluxetable}{lrrrrrrcrr}
\tabletypesize{\footnotesize}
\rotate
\tablewidth{0pt}
\tablecaption{Observed Targets\label{tab:opt}}
\tablehead{
\colhead{Name (SDSS J)} &
\colhead{$z_{em}$\tablenotemark{a}} &
\colhead{$i$} &
\colhead{$M_i$} &
\colhead{$\Delta v_{\rm b}$\tablenotemark{b}} &
\colhead{$\Delta(g-i)$} &
\colhead{$\alpha_{\rm o}$\tablenotemark{c}} &
\colhead{$E$$($$B$$-$$V$$)$} &
\colhead{\nh\tablenotemark{d}} &
\colhead{Sample\tablenotemark{e}}
}
\startdata
000654.10$-$001533.4 & 1.720 & 17.829 & $-27.18$ & $-$33$\pm28$  &
$0.156\pm0.030$ & $-0.09$ & 0.033 & 3.16 & S/I\\
015650.28$+$005308.4 & 1.652 & 18.239 & $-26.67$ & 11$\pm15$    &
$0.413\pm0.028$ & $-1.25$ & 0.029 & 2.69 & S/I\\
115115.38$+$003826.9 & 1.884 & 17.547 & $-27.67$ & 270$\pm28$   &
$0.060\pm0.024$ & $-0.46$ & 0.021 & 2.26 & S/II\\
005102.42$-$010244.3 & 1.890 & 17.367 & $-27.84$ & 1651$\pm34$  &
$-0.014\pm0.030$ & \phantom{$-$}0.12 & 0.041 & 2.83 & L/III\\
014812.23$+$000153.3 & 1.712 & 17.408 & $-27.59$ & 1980$\pm177$ &
$-0.145\pm0.028$ & $-0.02$ & 0.037 & 2.88 & L/IV\\
020845.54$+$002236.0 & 1.900 & 16.734 & $-28.49$ & 1738$\pm35$  &
$-0.023\pm0.035$ & $-0.43$ & 0.027 & 2.78 & L/IV\\
\tableline
\multicolumn{9}{c}{Archival Targets} \\
003131.44$+$003420.2 & 1.732 & 18.425 & $-26.62$ & 264$\pm68$ &
$0.122\pm0.038$ & $-0.24$ & 0.024 & 2.41 & S/II\\
173716.55$+$582839.4 & 1.775 & 18.595 & $-26.49$ & 433$\pm69$  &
$0.047\pm0.023$ & $-0.15$ & 0.042 & 3.51 & S/II\\
234819.58$+$005721.4 & 2.160 & 18.611 & $-26.93$ & 617$\pm136$ &
$-0.115\pm0.043$ & $-0.14$ & 0.025 & 3.81 & M/IND\\
011309.06$+$153553.5 & 1.809 & 17.839 & $-27.28$ & 705$\pm29$ &
$-0.042\pm0.029$ & \phantom{$-$}0.03 & 0.070 & 4.38 & M/IND\\
124540.99$-$002744.8 & 1.687 & 18.157 & $-26.83$ & 726$\pm1552$ &
$0.228\pm0.028$ & $-0.58$ & 0.024 & 1.73 & NA/NA\\
143841.95$+$034110.3 & 1.740 & 17.754 & $-27.27$ & 788$\pm36$  &
$0.313\pm0.022$ & $-1.20$ & 0.043 & 2.62 & M/IND\\
120436.63$+$015025.6 & 1.936 & 18.395 & $-26.88$ & 1247$\pm136$ &
$-0.099\pm0.030$ & $-0.84$ & 0.026 & 1.88 & L/III\\
020022.01$-$084512.1\tablenotemark{e} & 1.942 & 18.283 
& $-27.02$ & 337$\pm18$ & $0.398\pm0.033$ & $\cdots$\phantom{$-$} & 0.024 & 2.12 & NA\\
\enddata
\tablenotetext{a}{From \ion{Mg}{2}.}
\tablenotetext{b}{Velocity difference (in \kms) between the \CIV\ and
  \MgII\ emission-line redshifts.}
\tablenotetext{c}{The spectral index for a power-law fit to the
  UV continuum between rest-frame 1450 and 2200\,\AA.}
\tablenotetext{d}{The values for \nh\ (in units of 10$^{20}$\cmsq) used in the X-ray simulations
  are from NRAO Galactic \HI\ maps \citep{HI_ref}.}
\tablenotetext{e}{The sample labels indicate the membership of the
  quasar in the joint-spectral fitting samples.  Quasars with matching
  labels were fit together with $\Gamma$ and intrinsic \nh\ tied (\S\ref{sec:joint}).  
  Key: S, M, and L correspond to small-, moderate-, and large-blueshift,
  respectively;  I, II, III, and IV refer to smaller samples grouped by
  blueshift.  Those quasars marked ``IND''
  were fit individually, and ``NA'' indicates a quasar was not included.}
\tablenotetext{e}{This radio-loud quasar has broad absorption lines
  (see \S\ref{sec:bal}).}
\end{deluxetable}
\begin{deluxetable}{lccccrrrr}
\tabletypesize{\small}
\rotate
\tablewidth{0pt}
\tablecaption{Observing Log
\label{tab:log}
}
\tablehead{
\colhead{Name} &
\colhead{Obs. ID} &
\colhead{Date (MJD)} &
\colhead{$\theta$\tablenotemark{a}} &
\colhead{Exposure} &
\multicolumn{2}{c}{Counts\tablenotemark{b}} &
\colhead{Count Rate\tablenotemark{b}} &
\colhead{\HR\tablenotemark{c}} \\
\colhead{(SDSS J)} &
\colhead{} &
\colhead{} &
\colhead{(\arcmin)} &
\colhead{Time (ks)} &
\colhead{Soft} &
\colhead{Hard} &
\colhead{(10$^{-3}$~ct~\persec)} &
\colhead{} 
}
\startdata
000654.10$-$001533.4 & 4096 & 2003 Aug 02 (52853) &  0.0 & 4.45 & $34^{+6.9}_{-5.8}$      & $9^{+4.1}_{-2.9}$       &  $9.88^{+1.73}_{-1.48}$ &  $-0.58^{+0.15}_{-0.13}$ \\
015650.28$+$005308.4 & 4100 & 2003 Feb 23 (52693) &  0.0 & 5.57 & $55^{+8.5}_{-7.4}$      & $34^{+6.9}_{-5.8}$      & $15.97^{+1.88}_{-1.69}$ & $-0.24^{+0.11}_{-0.11}$  \\
115115.38$+$003826.9 & 4101 & 2003 Mar 03 (52701) &  0.0 & 3.69 & $38^{+7.2}_{-6.1}$      & $11^{+4.4}_{-3.3}$      & $13.29^{+2.18}_{-1.89}$ & $-0.55^{+0.14}_{-0.12}$  \\
005102.42$-$010244.3 & 4097 & 2002 Nov 20 (52598) &  0.0 & 3.50 & $42^{+7.5}_{-6.5}$      & $9^{+4.1}_{-2.9}$       & $14.56^{+2.34}_{-2.03}$ & $-0.65^{+0.13}_{-0.11}$  \\
014812.23$+$000153.3 & 4098 & 2003 Jun 25 (52815) &  0.0 & 3.71 & $24^{+6.0}_{-4.9}$      & $2^{+2.7}_{-1.3}$       &  $7.29^{+1.69}_{-1.39}$ & $-0.85^{+0.17}_{-0.10}$  \\
020845.54$+$002236.0 & 4099 & 2003 Aug 26 (52877) &  0.0 & 3.54 & $40^{+7.4}_{-6.3}$      & $11^{+4.4}_{-3.3}$      & $14.39^{+2.31}_{-2.01}$ & $-0.57^{+0.14}_{-0.12}$  \\
\tableline
\multicolumn{8}{c}{Archival Targets} \\
003131.44$+$003420.2 & 2101 & 2001 May 20 (52080) & 1.0 & 6.69 & $51.8^{+8.3}_{-7.2}$    & $9.4^{+4.3}_{-3.1}$     & $9.16^{+1.33}_{-1.17}$  & $-0.69^{+0.12}_{-0.10}$  \\ 
173716.55$+$582839.4 & 3038 & 2002 Aug 05 (52491) & 3.7 & 4.62 & $16.9^{+5.2}_{-4.1}$    & $<5.4$                  & $4.09^{+1.18}_{-0.94}$  &  $<-0.52$                \\
234819.58$+$005721.4 & 861  & 2000 May 26 (51691) & 0.0 & 74.20 & $285.9^{+18.0}_{-16.9}$ & $45.1^{+7.9}_{-6.8}$    & $4.46^{+0.26}_{-0.25}$  & $-0.73^{+0.04}_{-0.04}$  \\
011309.06$+$153553.5 & 3219 & 2002 Oct 18 (52565) & 6.2 & 58.50 & $390.0^{+20.9}_{-19.8}$ & $117.4^{+12.0}_{-11.0}$ & $8.67^{+0.41}_{-0.39}$  & $-0.54^{+0.04}_{-0.04}$  \\
124540.99$-$002744.8 & 4018 & 2003 Feb 14 (52684) & 8.1 & 4.94 & $64.4^{+9.1}_{-8.0}$    & $23.0^{+6.0}_{-4.9}$    & $17.69^{+2.12}_{-1.91}$ & $-0.47^{+0.11}_{-0.10}$  \\
143841.95$+$034110.3 & 3290 & 2002 Mar 13 (52347) & 6.9 & 57.57 & $218.8^{+15.9}_{-14.8}$ & $75.3^{+9.9}_{-8.9}$    & $5.11^{+0.32}_{-0.30}$  & $-0.49^{+0.06}_{ -0.05}$ \\ 
120436.63$+$015025.6 & 3234 & 2002 Nov 24 (52603) & 4.1 & 29.96 & $103.7^{+11.5}_{-10.4}$ & $28.2^{+6.6}_{-5.5}$    & $4.40^{+0.43}_{-0.39}$  & $-0.57^{+0.08}_{-0.08}$  \\
020022.01$-$084512.1 & 3265 & 2002 Oct 02 (52549) & 9.4 & 17.91 & $36.5^{+7.1}_{-6.1}$    & $15.6^{+5.2}_{-4.1}$    & $2.91^{+0.47}_{-0.41}$  & $-0.40^{+0.15}_{-0.14}$  \\
\enddata
\tablenotetext{a}{Angular distance in arcmin from the ACIS optical axis.}
\tablenotetext{b}{Detections for the full, soft, and hard bands are
  determined by {\em wavdetect}, and the background-subtracted counts are determined from aperture
  photometry (as described in \S\ref{sec:data}). Errors are 1$\sigma$ Poisson errors \citep{Gehrels}, except for
  non-detections where the limits are the $90\%$ confidence limits
  from Bayesian statistics \citep*{kbn}. The count rate is for the full band, 0.5--8.0~keV.}
\tablenotetext{c}{The \HR\ is defined as $(h-s)/(h+s)$, where $h$ and
  $s$ are the counts in the hard (2.0--8.0~keV) and soft (0.5--2.0~keV)
  bands, respectively. Errors in the \HR\ are propagated from the
  counting errors using the numerical method of \citet{Lyons1991}.}
 \end{deluxetable}
\begin{deluxetable}{lcccccccr}
\tabletypesize{\footnotesize}
\rotate
\tablewidth{0pt}
\tablecaption{X-ray Properties
\label{tab:xcalc}
}
\tablehead{
\colhead{Name (SDSS J)} &
\colhead{\GHR\tablenotemark{a}} &
\colhead{$\log(F_{\rm X})$\tablenotemark{b}} &
\colhead{$\log(f_{\rm 2 keV})$\tablenotemark{c}} &
\colhead{SDSS MJD} &
\colhead{$\log(f_{\rm 2500})$\tablenotemark{c}} &
\colhead{$\log(L_{\rm 2500})$\tablenotemark{d}} &
\colhead{\aox} &
\colhead{\daox\tablenotemark{e}} 
%
}
\startdata
000654.10$-$001533.4 & 1.85$^{+0.42}_{-0.35}$ & $-13.149\pm0.070$ &
$-30.910\pm0.070$ & 52519 & $-26.736\pm0.025$ & $31.132\pm0.025$ &
$-1.60\pm0.09$ & 0.06\\
015650.28$+$005308.4 & 1.08$^{+0.22}_{-0.20}$ & $-12.751\pm0.049$ &
$-30.955\pm0.049$ & 51871 & $-26.755\pm0.018$ & $31.079\pm0.018$ &
$-1.61\pm0.09$ & 0.05\\
115115.38$+$003826.9 & 1.75$^{+0.35}_{-0.33}$ & $-13.012\pm0.067$ &
$-30.803\pm0.067$ & 51943 & $-26.626\pm0.014$ & $31.311\pm0.014$ &
$-1.60\pm0.09$ & 0.08\\
005102.42$-$010244.3 & 2.03$^{+0.37}_{-0.36}$ & $-13.033\pm0.065$ &
$-30.685\pm0.065$ & 52201 & $-26.517\pm0.016$ & $31.422\pm0.016$ &
$-1.60\pm0.08$ & 0.09\\
014812.23$+$000153.3 & 2.90$^{+0.90}_{-0.77}$ & $-13.437\pm0.092$ &
$-30.849\pm0.092$ & 52199 & $-26.554\pm0.015$ & $31.309\pm0.015$ &
$-1.65\pm0.11$ & 0.03\\
020845.54$+$002236.0 & 1.82$^{+0.34}_{-0.34}$ & $-12.984\pm0.065$ &
$-30.734\pm0.065$ & 52178 & $-26.210\pm0.014$ & $31.734\pm0.014$ &
$-1.74\pm0.09$ & $-0.02$\\
\tableline
\multicolumn{7}{c}{Archival Targets} \\
003131.44$+$003420.2 & $1.95^{+0.35}_{-0.35}$ & $-13.294\pm0.059$ &
$-31.009\pm0.059$ & 51793 & $-26.711\pm0.025$ & $31.163\pm0.025$ &
$-1.65\pm0.08$ & 0.02\\
173716.55$+$582839.4 & $>1.45$                & $-13.270\pm0.113$ &
$-30.951\pm0.113$ & 52017 & $-26.890\pm0.040$ & $31.001\pm0.040$ &
$-1.56\pm0.14$ & 0.09\\
234819.58$+$005721.4 & $2.07^{+0.17}_{-0.13}$ & $-13.642\pm0.025$ &
$-31.237\pm0.025$ & 51788 & $-26.968\pm0.086$ & $31.076\pm0.086$ &
$-1.64\pm0.04$ & 0.02\\
011309.06$+$153553.5 & $1.88^{+0.10}_{-0.10}$ & $-13.092\pm0.020$ &
$-30.830\pm0.020$ & 51878 & $-27.039\pm0.019$ & $30.865\pm0.019$ &
$-1.45\pm0.07$ & 0.19\\
124540.99$-$002744.8 & $1.62^{+0.25}_{-0.25}$ & $-12.727\pm0.050$ &
$-30.605\pm0.050$ & 51928 & $-27.099\pm0.020$ & $31.141\pm0.020$ &
$-1.35\pm0.09$ & 0.28\\
143841.95$+$034110.3 & $1.73^{+0.12}_{-0.15}$ & $-13.299\pm0.026$ &
$-31.120\pm0.026$ & 52023 & $-26.734\pm0.017$ & $31.141\pm0.017$ &
$-1.68\pm0.06$ & $-0.02$\\
120436.63$+$015025.6 & $1.57^{+0.25}_{-0.25}$ & $-13.503\pm0.040$ &
$-31.387\pm0.040$ & 52017 & $-26.750\pm0.101$ & $31.207\pm0.101$ &
$-1.78\pm0.08$ & $-0.11$\\
020022.01$-$084512.1 & $1.48^{+0.32}_{-0.30}$ & $-13.464\pm0.066$ &
$-31.404\pm0.066$ & 52149 & $-26.833\pm0.020$ & $31.129\pm0.020$ &
$-1.75\pm0.09$ & $-0.09$\\
\enddata
\tablenotetext{a}{\GHR\ is a coarse measure of the hardness of the X-ray
  spectrum determined by comparing the observed \HR\ (see
  Table~\ref{tab:log}) to a simulated \HR\ that takes into account
  spatial and temporal variations in the instrument response (see
  $\S$\ref{sec:data}.2).}
\tablenotetext{b}{The full-band X-ray flux, $F_{\rm X}$, has units of \flux\ 
  and is calculated by integrating the power-law spectrum given by $\Gamma$
  and normalized by the full-band count rate from 0.5--8.0~keV.  The errors
  are derived from the 1$\sigma$ errors in the full-band count rate.}
\tablenotetext{c}{X-ray and optical flux densities were measured at
  rest-frame 2~keV and 2500\,\AA, respectively; units are \fnu.}
\tablenotetext{d}{The 2500\,\AA\ luminosity density, $L_{\rm 2500}$, has units of \lumin~Hz$^{-1}$.}
\tablenotetext{e}{The parameter \daox\ is the difference between the observed \aox\ and the expected \aox\
  calculated from Equation 4 of \citet{Vignali2003} using the luminosity density at 2500\,\AA.}
\end{deluxetable}
\begin{deluxetable}{lcccc}
\tabletypesize{\small}
\tablewidth{0pt}
\tablecaption{Results from Joint-Spectral Fitting
\label{tab:spec}
}
\tablehead{
\colhead{Sample\tablenotemark{a}} &
\colhead{$\Gamma$\tablenotemark{b}} &
\colhead{\nh\tablenotemark{b}} &
\colhead{$C$-stat/$\nu$} &
\colhead{Total} \\
\colhead{} &
\colhead{} &
\colhead{(10$^{22}$\cmsq)} &
\colhead{} &
\colhead{Counts} 
}
\startdata
1. Large -- New (3) & $2.00^{+0.61}_{-0.62}$ & $1.56^{+1.87}_{-1.53}$ &
426.8/1536 & 129 \\
2. Small -- New (3)& $1.42^{+0.35}_{-0.26}$ & $0.00^{+0.86}_{-0.00}$ &
553.2/1536 & 180 \\
3. Large -- New (3) $+$ Archival (1) & $1.98^{+0.82}_{-0.29}$ & $1.47^{+2.27}_{-0.93}$ &
648.2/2048 & 266 \\
4. Small -- New (3) $+$ Archival (2) & $1.64^{+0.32}_{-0.20}$ & $0.00^{+0.63}_{-0.00}$ &
801.8/2560 & 260 \\
5. Moderate -- Archival (3) & $1.98^{+0.31}_{-0.13}$ & $0.27^{+0.86}_{-0.27}$ &
996.9/1536 & 1139\phantom{1}\\
\enddata
\tablenotetext{a}{The samples are described in more detail in
  $\S$\ref{sec:joint}.2.  The numbers in parentheses refer to the
  number of quasars in each sample.}
\tablenotetext{b}{The errors quoted are for 90$\%$ confidence ($\Delta
  C = 4.61$ for two parameters of interest).  Both $\Gamma$ and
\nh\ are tied together to determine the average parameter values for each
sample.  The redshift and Galactic \nh\ for each quasar are fixed to
the appropriate values (see Table~\ref{tab:opt}).}
\end{deluxetable}
\end{document}